\documentclass[11pt,preprint,preprintnumbers,a4paper]{article}
\pdfoutput=1

\usepackage{jheppub}

\usepackage{afterpage}

\usepackage{longtable}
\usepackage{multirow}
\usepackage{enumerate}
\usepackage{amsmath}
\usepackage{amsfonts}
\usepackage{amssymb}
\usepackage{graphicx, rotating}
\usepackage{epstopdf}
\usepackage{epsfig}
\usepackage{latexsym}
\usepackage{graphicx}
\usepackage{color}
\usepackage{amsmath,amssymb}
\usepackage{slashed}
\usepackage{hyperref}
\usepackage{wasysym}
\usepackage{MnSymbol}
\usepackage{multirow}
\usepackage{caption}
\usepackage{slashed}

\definecolor{LightCyan}{rgb}{0.88,1,1}
\definecolor{LightViolet}{rgb}{1, 0.88,1}
\definecolor{LightGreen}{rgb}{0.88, 1, 0.88}
\definecolor{forestgreen}{rgb}{0.13,0.35,0.13}

\definecolor{oucrimsonred}{rgb}{0.6, 0.0, 0.0}
\definecolor{persianblue}{rgb}{0.11, 0.22, 0.73}
\definecolor{forestgreen}{rgb}{0.13,0.35,0.13}
 \hypersetup{colorlinks, citecolor=oucrimsonred, linkcolor=persianblue, urlcolor=oucrimsonred}

\usepackage{color, colortbl}
\definecolor{rossos}{cmyk}{0,1,1,0.55}
\definecolor{bluscuro}{rgb}{0.15, 0.2, .85}
\definecolor{bluchiaro}{cmyk}{1,.3,0.,0.1}
\definecolor{verdechiaro}{rgb}{0.6,.1,0.9}
\definecolor{forestgreen}{rgb}{0.13,0.35,0.13}
\definecolor{gold}{rgb}{1,0.84,0}
\definecolor{darkcyan}{rgb}{0,0.55,0.55}
\definecolor{brown}{rgb}{0.65,0.16,0.16}
\definecolor{orange}{rgb}{1,0.65,0.}

\def\bea{\begin{eqnarray}} \def\eea{\end{eqnarray}}
\def\be{\begin{equation}} \def\ee{\end{equation}}

\def\cO{{\cal O}}
\def\met{\slashed{E}_T}
\newcommand{\fb}{{\rm fb}}
\newcommand{\br}{{\rm BR}}
\newcommand{\beq}{\begin{equation}}
\newcommand{\eeq}{\end{equation}}
\newcommand{\cL}{{\cal L}}
\newcommand{\cP}{{\cal P}}

\newcommand{\promille}{%
  \relax\ifmmode\promillezeichen
        \else\leavevmode\(\mathsurround=0pt\promillezeichen\)\fi}
\newcommand{\promillezeichen}{%
  \kern-.05em%
  \raise.5ex\hbox{\the\scriptfont0 0}%
  \kern-.15em/\kern-.15em%
  \lower.25ex\hbox{\the\scriptfont0 00}}

\begin{document}
\preprint{CERN-TH-2016-035}

\vspace*{-30mm}

\title{\boldmath Precision Drell-Yan Measurements at the LHC and Implications for the Diphoton Excess}

\author[a]{Florian Goertz,}
\author[a,b]{Andrey Katz,}
\author[a,c]{Minho Son,}
\author[a]{and Alfredo Urbano}

\affiliation[a]{Theoretical Physics Department, CERN, Geneva, Switzerland}
\affiliation[b]{Universit\'{e} de Gen\`{e}ve, Department of Theoretical Physics and Center for Astroparticle Physics,\\
24 quai E. Ansermet, CH-1211, Geneva 4, Switzerland}
\affiliation[c]{ Department of Physics, Korea Advanced Institute of Science and Technology,\\
335 Gwahak-ro, Yuseong-gu, Daejeon 305-701, Korea}

\emailAdd{florian.goertz@cern.ch}
\emailAdd{andrey.katz@cern.ch}
\emailAdd{minho.son@kaist.ac.kr}
\emailAdd{alfredo.leonardo.urbano@cern.ch}

\vspace{2cm}


\abstract{
Precision measurements of the Drell-Yan (DY) cross sections at the LHC constrain new physics scenarios 
that involve new states with electroweak (EW) charges. 
We analyze these constraints and apply 
them to models that can address the LHC diphoton excess 
at 750~GeV. We confront these findings with LEP 
EW precision tests and  show that DY  provides
stronger constraints  than the  LEP data.
While 8~TeV data can already probe some parts of the interesting region of parameter space,
LHC14 results are expected to cover a substantial part of the relevant terrain. 
We derive the bounds from the existing data, estimate LHC14 reach and compare them to the 
bounds one gets from LEP and future FCC-ee precision measurements.}  

\maketitle

\section{Introduction}\label{sec:Intro}

Recently both ATLAS and CMS reported an excess in the search for diphoton resonances around 
$M_{\gamma \gamma} \sim 750$~GeV with 3.2~fb$^{-1}$ and 2.6~fb$^{-1}$ of integrated luminosity at 
$\sqrt s = 13$~TeV, respectively.
The local (global) significance of the excess is 3.9~\!$\sigma$ (2.3~\!$\sigma$) for the ATLAS data, 
with the best-fit value for the width of the resonance $\Gamma \sim 45$~GeV~\cite{ATLAS-CONF-NOTE}. 
CMS reported a local excess with significance of  2.6~\!$\sigma$ at a mass compatible with ATLAS, 
assuming a narrow width. 
This significance goes down to 2~\!$\sigma$ if $\Gamma = 45$~GeV is assumed~\cite{CMS:2015dxe}. 
The findings are compatible with searches at $\sqrt s = 8$~TeV, given that the production cross section
of the potential resonance $S$  
increases by about a factor of 5 for the larger center-of-mass energy.
This is, for example, realized if the resonance with the mass around 750~GeV
is produced in gluon-gluon or $b\bar b$ fusion~\cite{Franceschini:2015kwy}. Although these hints are by no 
means decisive and it is possible, that the origin of both is in somewhat unlikely fluctuations of the background, 
it is very interesting to understand the consequences of interpreting this excess as a true new physics resonance.

The tentative large width, however, poses challenges for model building, hinting to a large
number of new states with sizable charges mediating the decay $S \to \gamma \gamma$ in a
weakly coupled framework~\cite{Knapen:2015dap,Ellis:2015oso,McDermott:2015sck,Angelescu:2015uiz,Gupta:2015zzs,Bauer:2015boy}. 
This width implies  that also the partial width into photons 
should be sizable, at least of order 
 $2 \times 10^{-4} \ M_S$, because there are considerable constraints on the relative size of other possible
 decay channels of $S$ from the 8~TeV data.

The most popular and reasonable realization of the large number of the new states would be vector-like fermions, charged
under the electroweak (EW) force.\footnote{Any other solution could potentially introduce difficult conceptual  problems, 
like flavor-changing
neutral currents in the case of chiral representations under the SM.  Scalars will also have difficulties, 
which were discussed in detail in Ref.~\cite{Salvio:2016hnf}. We 
comment more on the scalar case in Section~\ref{sec:scalars}.}  Since these 
new states have EW production cross section (and, possibly, very difficult decay 
modes for experimental detection) they can relatively easily evade the direct LHC searches and still be sufficiently light, well below the TeV scale.  
It has been noticed in 
several works before, that these new vector like states significantly modify the running of the hypercharge 
coupling~\cite{Gu:2015lxj, Goertz:2015nkp, Son:2015vfl}. This in turn leads to new constraints, originating from the consistency 
of these models, perturbativity and the scale of the 
Landau pole, which all have been discussed in detail in the above mentioned references.

Interestingly, on top of these constraints, one can put \emph{experimental constraints} on the existence of such 
large number of new vector-like EW states. 
As we will show in detail, the presence of such vector-like fermions can be indirectly tested
by measuring the neutral Drell-Yan (DY) process at the LHC, considering their impact on the running of the hypercharge coupling. We 
will demonstrate that large portions of the parameter space, relevant for the 750~GeV diphoton resonance,
can be probed in a relatively model independent way via the LHC DY production far away from the $Z$-pole.

The idea to explore the running of the EW couplings to probe new EW states 
was carefully elaborated on in Ref.~\cite{Alves:2014cda}.
This paper has shown that the running of EW couplings, $\alpha_2$ and $\alpha_Y$, can be successfully probed 
at the LHC14 and at a future 100~TeV collider, putting new limits that are much stronger than one gets from LEP measurements. 
For
example, it is claimed that the high-luminosity LHC will be sensitive to deviations in $\alpha_2$ of less than 10\% 
from the SM value at a scale of 2.5~TeV. In this work we take this idea one step further, and show that precisely the same 
measurements at LHC8 already put meaningful constraints on the models that are explaining the 750~GeV resonance. 
We show, for example, that for the EW states around 400~GeV, values of $NQ^2 \sim 60$ 
(with $N$ being a total multiplicity 
number and $Q$ the hypercharge)  are in tension with the DY data, already reducing the parameter space 
for the large width interpretation of the 750~GeV resonance. Future measurements at LHC14 will further shed light 
on the parameter space of the $S$, which will be particularly powerful in the latter case. 

Our paper in organized as follow. In Section~\ref{sec:EWPT} we present the perturbative model, which might address the 750~GeV LHC diphoton excess. We also discuss the necessity of the new EW states, explain how they affect the LEP measurements (via changing the Y-parameter) and show the LEP bounds. We will see that these bounds are fairly weak. In Section~\ref{sec:DY}
we briefly overview the theory of DY production at the LHC, with an emphasis on the possible change in the cross sections due 
to new physics which affects the hypercharge coupling running.
In Section~\ref{sec:Analysis} we describe our statistical procedure, show the bounds that we get from LHC8 and provide the sensitivity 
projection for LHC14. Here, we also discuss NLO, PDF and other systematic uncertainties.
We summarize the implications of our findings on the 750~GeV diphoton resonance and 
discuss some future developments in Section~\ref{sec:Future}. 
Finally, we briefly comment on direct searches for the new EW states in Section~\ref{sec:DS} and conclude. The technical details concerning the simulation of the SM prediction for the DY process are
relegated to Appendix~\ref{sec:AppA}.

\section{Toy Model for 750~GeV Excess and EW Precision Tests}
\label{sec:EWPT}

\subsection{Overview of data and interpretation}
If we interpret the LHC data as a signal for a new resonance in the $\gamma \gamma$ channel, the suggested 
production cross sections are~\cite{Franceschini:2015kwy} 
\beq
\sigma (pp \to S ) \times {\br} (S \to \gamma \gamma) = (10 \pm 3)~\fb \ \ / \ \  (6 \pm 3)~\fb
\eeq 
at ATLAS~/~CMS, respectively. ATLAS data prefers a relatively wide resonance, $\Gamma_S/M_S \approx 0.06$. 
It is important to mention that CMS data does not prefer large width, and it is unclear whether the wide resonance 
assumptions notably improves the \emph{overall} fit, when the LHC8 and LHC13 data of both experiments is taken into account. For example, 
it is claimed in Ref.~\cite{Buckley:2016mbr} that narrow width of the resonance is more likely if the Run~II information of both ATLAS and CMS is considered, while the preference to large width is merely marginal, when it is combined with the Run~I 
data.\footnote{For more discussions on the width of the new resonance and its possible consequences see 
e.g.~\cite{Knapen:2015dap,Gupta:2015zzs,Falkowski:2015swt}.  } 

For concreteness we consider the following effective theory for $S$, assuming that $S$ is a scalar and its couplings 
do not violate CP:
\beq\label{eq:effcoup}
\cL_{eff} = \frac{e^2 S F_{\mu \nu} F^{\mu \nu}}{2\Lambda_\gamma} + 
\frac{g_3^2 S G_{\mu \nu}G^{\mu \nu} }{2 \Lambda_g}~.
\eeq
If $S$ is a pseudoscalar, one essentially gets the same couplings with the obvious replacements 
$F_{\mu \nu} F^{\mu \nu} \to F_{\mu \nu} \tilde F^{\mu \nu}$ and  
$G_{\mu \nu} G^{\mu \nu} \to G_{\mu \nu} \tilde G^{\mu \nu}$. Here we also assumed that the dominant 
production channel for $S$ is via gluon fusion. Of course this is not the only option, and production from 
heavy flavors can be even slightly favored by data if LHC8 is taken into account~\cite{Franceschini:2015kwy}. 

In a perturbative model, the couplings in Eq.~\eqref{eq:effcoup} are induced by new states, charged under the EM and strong force, respectively. 
As we have alluded, we will further study the model, in which these couplings are induced by new vector-like fermions. 
In principle, one can induce the coupling of $S$ to photons either via introducing new vector like representations 
under $SU(2)_L$, or under hypercharge (or both). 
For simplicity, we will focus on the latter case 
and consider $N_X$ new vector-like fermions $X_i,$\,{\small $i=1,..,N_X$}, only charged under hypercharge. Moreover, we assume that they all share the same mass $M_{X_i} \equiv M_X$, quantum numbers, and couplings. In this case, one easily 
matches the scale $\Lambda_\gamma$ as\footnote{We will later introduce $d_X$ additional degrees of freedom
for each vector-like fermion $X_i$, describing for example a color charge (with, {\it e.g.}, $d_X=3$ the dimension 
of the color representation), leading to the 
replacement $N_X \to d_X N_X$. With this, they 
could simultaneously generate $\Lambda_g < \infty$, however in this article we will be agnostic and not make any assumption on the UV physics inducing the operator relevant for $S$ {\it production}.}
\beq\label{eq:matching}
\Lambda_\gamma^2 = \frac{16 \pi^4 M_S^2}{(N_XQ_X^2)^2 y^2 \tau_X |\mathcal{S}(\tau_X)|^2}\,, \ \text{with }
\Gamma(S \to \gamma \gamma) = \frac{M_S^3}{80 \pi \Lambda_\gamma^2}\,.
\eeq  
Here, $Q_X$ denotes the common hypercharge of the $X_i$, 
$\tau_X \equiv 4M_X^2/M_S^2$ and, assuming $S$ is a scalar particle, 
\beq
\mathcal{S}(\tau) \equiv 1+ (1-\tau)\arctan^2\left(\frac{1}{\sqrt{\tau - 1}}\right)~,
\eeq 
while $y$ is the Yukawa coupling between the new scalar resonance and the vector-like states, $\cL \supset y S X X$. 
The pseudo-scalar case corresponds to the replacement  $\mathcal{S}(\tau)  \to \cP (\tau) \equiv \arctan^2\left(1/\sqrt{\tau -1}\right)$ together with trading $y$ for the pseudo-scalar Yukawa coupling.

What should the number $N_X Q_X^2$ be to match the data? This again depends on our assumptions. If we assume 
that $S$ is a narrow resonance, one can end up with a 
fairly small number of exotic vector-like fermions, which will have a relatively 
limited impact on the DY precision measurements. In this case the only constraint that the data imposes is 
\beq
\frac{\Gamma_{gg} \Gamma_{\gamma \gamma}}{M_S \Gamma} \sim 10^{-6}~,
\eeq
where we use the short-hand notations, $\Gamma_{\gamma\gamma} \equiv \Gamma(S\rightarrow \gamma\gamma), \Gamma_{gg}\equiv \Gamma(S\rightarrow gg)$.

Dijet searches constrain the width into gluons relatively weakly compared to the observed width into photons,
$\Gamma_{gg} \lesssim 1200\ \Gamma_{\gamma \gamma}$. This essentially means that if we assume narrow width, we can saturate $\Gamma\simeq\Gamma_{gg}$ and it can be sufficient to reproduce $\Gamma_{\gamma \gamma}/M_S \sim 10^{-6}$. 
This goal is easy to achieve with only $N_XQ_X^2 \sim \cO(5-10)$.

The situation dramatically changes if we try to reproduce the 6\% width of the resonance, which is preferred by ATLAS.   
As the width (into some other 
particles) grows substantially, one needs to increase the width to photons and is easily driven to $\Gamma_{\gamma \gamma}/M_S \gtrsim 10^{-3}$, 
depending on the dominant decay products $f$ of $S$, due to 8\,TeV upper bounds on $\Gamma(S \to f)/\Gamma_{\gamma \gamma}$, scaled to 13\,TeV \cite{Franceschini:2015kwy}. In any case, in the wide width scenario, the generic constraint from
8\,TeV data, $\Gamma \lesssim 1500 \times \Gamma_{\gamma\gamma}$ (considering standard decay modes), suggests that $\Gamma_{\gamma \gamma}/M_S$ cannot be lower than $\sim 10^{-4}$.
It is important to mention that this requirement is not unique for gluon fusion 
production and one gets roughly the same requirements of the partial width into diphotons if heavy flavor production is assumed.

The goal $\Gamma_{\gamma \gamma}/M_S \sim 10^{-3}$ is notoriously hard to achieve and it demands 
sufficiently large $N_X Q_X^2 \sim \cO(100-500)$, depending on the masses of the new fermions. These numbers are 
already big enough to significantly deflect the running of $\alpha_Y$ from the SM trajectory, such that 
the effects are clearly visible in DY production.

\subsection{LEP constraints}\label{sec:Lepconstraints}
Before analyzing in detail the bounds that we get from the LHC, we first show the bounds from LEP.
The $SU(2)_L$-singlet vector-like fermions $X$ with non-zero hypercharge $Y \equiv Q_X$ and mass $M_X$ contribute to the two-point function, $\Pi^{\mu\nu}_{BB}(q^2)$. Below the mass scale of the heavy fermions, it generates an effective operator $(\partial_\mu B^{\mu \nu})^2 $ which maps on the $\mathcal{Y}$ parameter~\cite{Barbieri:2004qk},
\begin{equation}
\mathcal{Y} = d_X N_X Q_X^2\, \frac{\alpha_Y m_W^2}{15\pi M_X^2}~,
\end{equation}
where $d_X N_X$ denotes the total number of degrees of freedom. The one-loop $\beta$ function\footnote{We use the hypercharge in the so-called GUT normalization $g_1 \equiv \sqrt{5/3} g_Y$.} of the hypercharge coupling is also encoded in the same two-point function, leading to the same parametric dependence $d_X N_X Q_X^2$, and is given by
\begin{equation}\label{eq:Beta} 
\beta_{g_1}^{(1)} = \left(
\frac{41}{10} + d_X N_X Q_X^2 \frac{4}{5}
\right) g_1^3 = 16\pi^2\frac{dg_1}{d\ln\mu}~,
\end{equation}
where the modification due to the new vector-like states alters the SM running above the threshold $\mu > M_X$.
Below the threshold, the heavy fermions $X$ are integrated out and the resulting one-loop $\beta$ function reduces to the SM one.

\begin{figure}[!htb!]
\minipage{0.5\textwidth}
  \includegraphics[width=.9\linewidth]{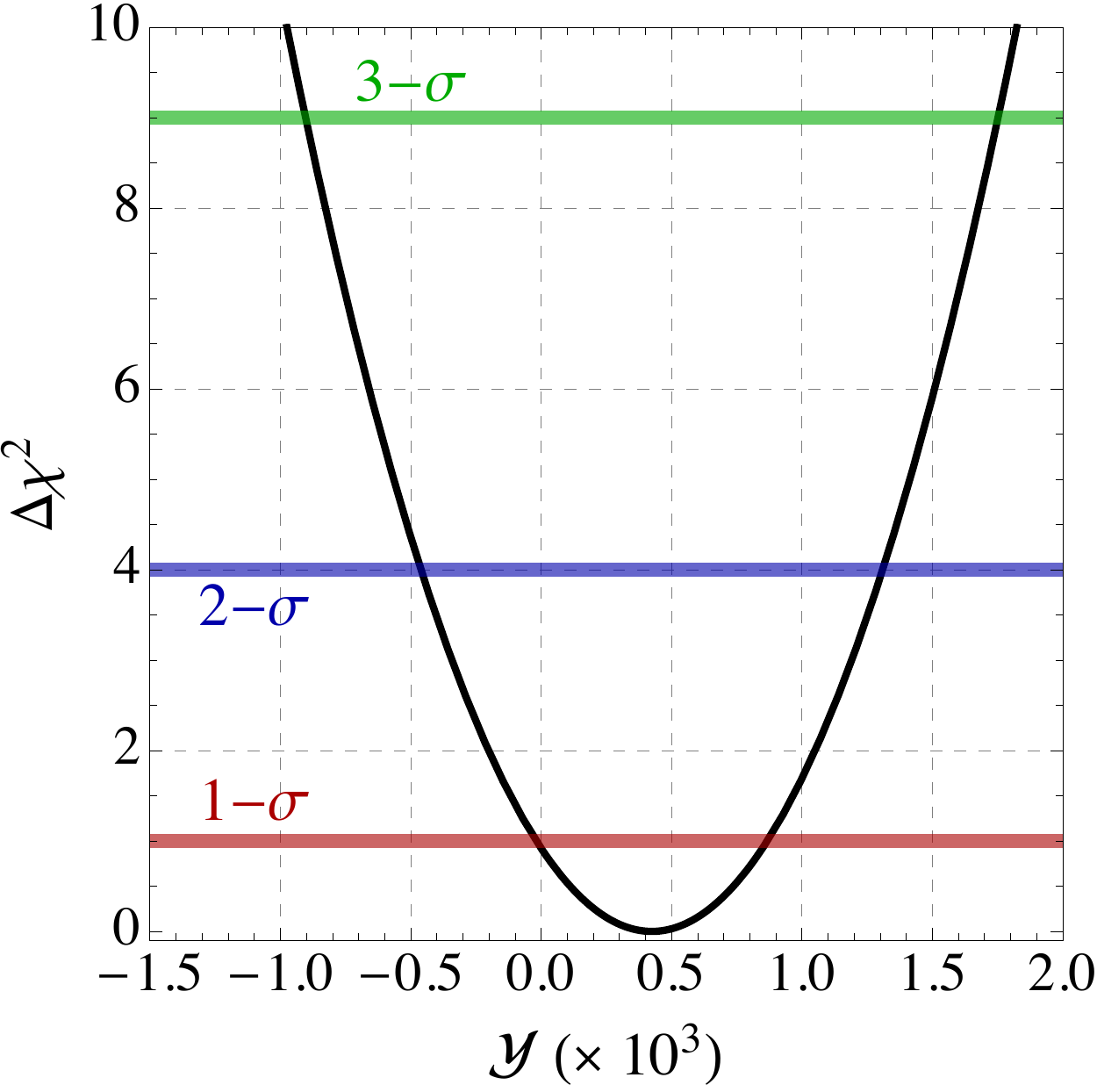}
\endminipage\hfill
\minipage{0.5\textwidth}
  \includegraphics[width=1.\linewidth]{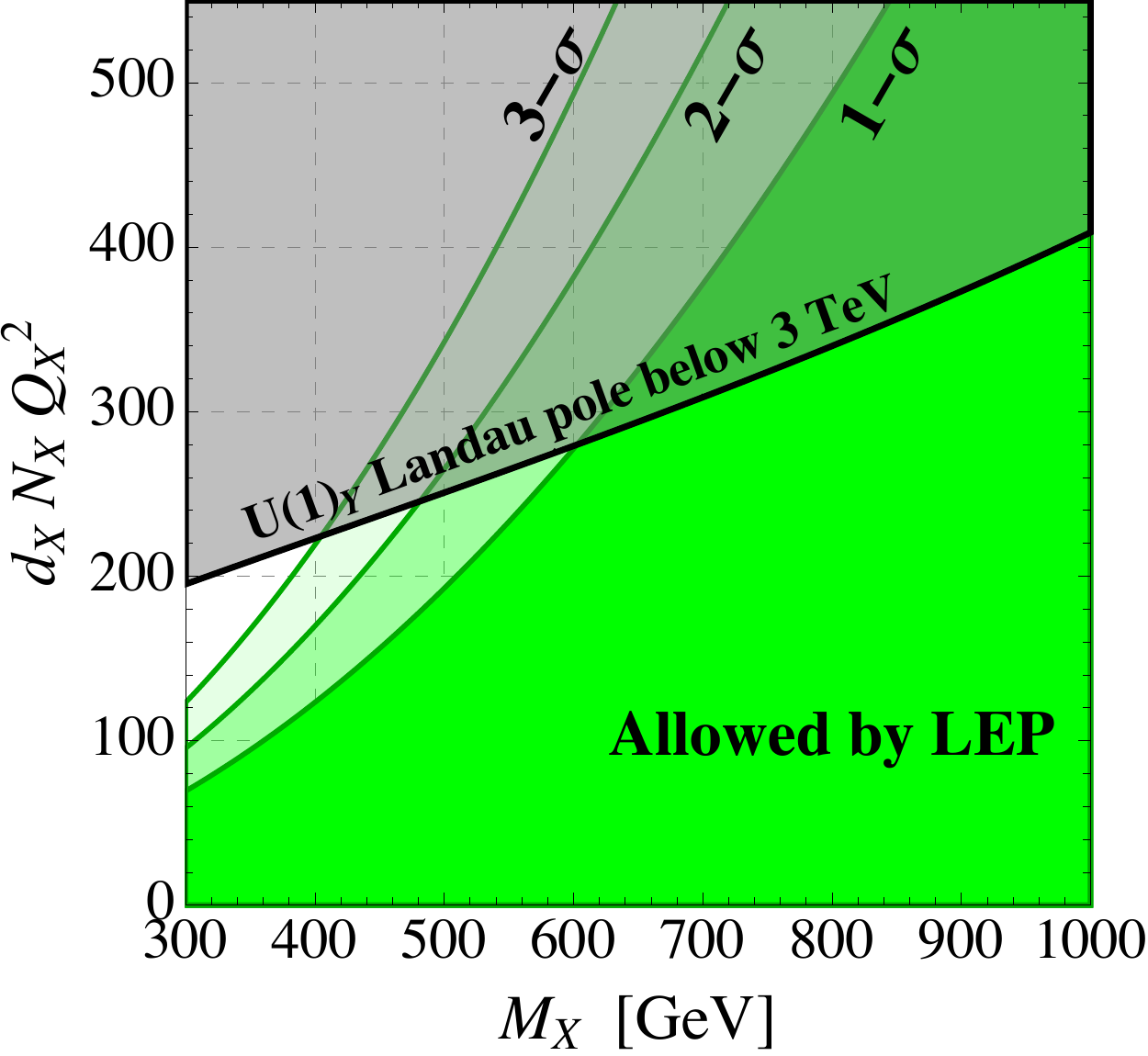}
\endminipage
\caption{\em 
Left panel. One-parameter fit of LEP data in terms of the $\mathcal{Y}$ oblique parameter. 
All the remaining oblique parameters are set to zero.
Right panel. Two-parameter fit of LEP data (allowed region in green) in term of the mass $M_{X}$ of the new vector-like states 
and the combination $d_X N_X Q_X^2$, where $d_X$ is the number of degrees of freedom (e.g. $d_X = 3$ for a color charge).
We superimpose (region shaded in gray) 
the parameter space in which the $U(1)_Y$ Landau pole is lowered below $3$ TeV~\cite{Son:2015vfl}.
}
\label{fig:EWPT}
\end{figure}

The best-fit of the $\mathcal{Y}$ parameter from LEP data and the recasted bounds on new vector-like fermions are illustrated in Fig~\ref{fig:EWPT}. As one would naively expect, these bounds are very weak.

\section{Drell-Yan at the LHC}\label{sec:DY}

In this section we show how one can estimate the DY double differential cross section 
in the presence of the $\alpha_Y$ running due to the new physics. 
The computation of the DY cross section at the LHC proceeds in the following steps.

\begin{itemize}

\item[$\circ$]  At the parton level and in the center of mass partonic reference frame
 the DY scattering process $q\bar{q} \to Z^*/\gamma \to l^+l^-$ is described by the following double-differential cross-section
\begin{equation}\label{eq:Partonic}
\left. \frac{1}{2\, M_{ll}} \frac{d^2\hat{\sigma}}{dz dM_{ll}}\right|_{q\bar{q}} = a^2 f_{q\bar{q}}(s_W^2, M_{ll},z) \times \frac{\pi M_{ll}^2}{2}\frac{\delta(\hat{s} - M_{ll}^2)}{(M_{ll}^2 - m_Z^2)^2 + \Gamma_Z^2m_Z^2}~,
\end{equation}
where $z \equiv \cos\theta$, being $\theta$ is the scattering angle in the partonic center of mass frame. 
The factor $a$ and the rescaling function in Eq.~(\ref{eq:Partonic}), following~\cite{Alves:2014cda}, are defined as
\begin{equation}\label{eq:Factor}
\begin{split}
a \equiv&\  \frac{\alpha}{s_W^2 c_W^2}~,\\
f_{q\bar{q}}(s_W^2, M_{ll},z)  \equiv&\ c_{q\bar{q}}^{(0)} + c_{q\bar{q}}^{(1)} s_W^2 + c_{q\bar{q}}^{(2)} s_W^4 + c_{q\bar{q}}^{(3)} s_W^6 + c_{q\bar{q}}^{(4)} s_W^8~.
\end{split}
\end{equation}
The dependence of the scattering cross-section on the running gauge couplings is encoded in $a^2 f_{q\bar{q}}(s_W^2, M_{ll}, z)$ by means of the relations
\begin{equation}\label{eq:sinW}
s_W^2 = \frac{g_Y^2}{g_Y^2 + g_L^2}~,~~~\alpha = \frac{g_L^2 s_W^2}{4\pi}~,
\end{equation}
where $g_L$ is the $SU(2)_L$ gauge coupling with one-loop $\beta$ function
$\beta^{(1)}_{g_L} = -(19/6)g_L^3$.

At this stage it is important to stress that the cross-section in 
Eq.~(\ref{eq:Partonic}) is derived in full generality with respect to the running of 
the gauge couplings $g_Y$ and $g_L$.
Since the energy scale of the process is set by $\sqrt{\hat{s}} = M_{ll}$, 
one has to use the running couplings evaluated at $\mu = M_{ll}$.
 The impact of new vector-like states with mass $M_X$  is captured 
 by solving Eq.~(\ref{eq:Beta}) with specific values of $d_X$, $N_X$, $Q_X$, 
and using the corresponding expression for $g_Y(\mu = M_{ll})$ in Eqs.~(\ref{eq:Partonic}-\ref{eq:sinW}). 
The SM prediction corresponds to $N_X = 0$.

The coefficients $c_{q\bar{q}}^{(i)}$, $i=0,\dots,4$ depend on dilepton invariant mass, quantum numbers of the initial quark-antiquark pair, and scattering angle.
We find\footnote{Notice that, comparing our results with~\cite{Alves:2014cda}, we found a factor $3$ discrepancy in the explicit expression of the coefficients $c_{q\bar{q}}^{(i)}$.}
\begin{eqnarray}
c_{q\bar{q}}^{(0)} &=& \frac{T_{3L}^2(1+z)^2}{48}~,\\
c_{q\bar{q}}^{(1)} &=& \frac{T_{3L}(1+z)^2}{24}\left[
(Q_q - 2T_{3L}) - 2Q_q\frac{m_Z^2}{M_{ll}^2}
\right]~,\\
c_{q\bar{q}}^{(2)} &=&  \frac{1}{24}\left\{
5Q_q^2(1+z^2) + 4T_{3L}^2(1+z^2) -2Q_qT_{3L}[3+z(-2+3z)]
\right\} \\
&+& \left\{
-6Q_q(1+z^2) + T_{3L}[5+ z(2 + 5z)]
\right\}\frac{Q_q m_Z^2}{12M_{ll}^2}
+ \frac{Q_q^2(1+z^2)}{3}\frac{m_Z^2(m_Z^2 + \Gamma_Z^2)}{M_{ll}^4}~,\nonumber \\
c_{q\bar{q}}^{(3)} &=&  \frac{Q(1+z^2)}{6}\left[
(3Q_q - 2T_{3L})\frac{m_Z^2}{M_{ll}^2} - 4Q_q \frac{m_Z^2(m_Z^2 - \Gamma_Z^2)}{M_{ll}^4}
\right]~,\\
c_{q\bar{q}}^{(4)} &=& \frac{Q_q^2(1+z^2)}{3}\frac{m_Z^2(m_Z^2 + \Gamma_Z^2)}{M_{ll}^4}~,
\end{eqnarray}
where $Q_{u} = 2/3$, $T_{3L} = 1/2$ ($Q_{d} = -1/3$, $T_{3L} = -1/2$) for, respectively, up- and down-type quarks.
At large dilepton invariant mass ($M_{ll} \gg m_Z$) $c_{q\bar{q}}^{(3,4)}\to 0$.
The scattering cross-section in Eq.~(\ref{eq:Partonic}) is averaged over the initial spin and color.
At the weak scale we use the numerical matching values $g_L(\mu = m_Z) = 0.649$, $g_1(\mu = m_Z) = 0.459$.

\item[$\circ$] The hadronic differential cross-section is obtained by convoluting with the PDFs
\begin{equation}
\frac{1}{2\, M_{ll}} \frac{d\sigma}{dM_{ll}} = \int_{0}^1 dx_1 dx_2 dz \sum_{q}
\left[f_q(x_1)f_{\bar{q}}(x_2) +f_{\bar{q}}(x_1)f_{q}(x_2) 
\right]
\left.\frac{d^2\hat{\sigma}}{dz dM_{ll}^2}\right|_{q\bar{q}}~,
\end{equation}
with $\hat{s} = x_1 x_2 s$, being $\sqrt{s}$ the total energy in the hadronic  center of mass frame. We introduce the variable $\tau \equiv x_1x_2$ which corresponds to the fraction of the energy transferred to the partonic system.
The rapidity of the lepton is defined as
\begin{equation}
{\rm y} \equiv \frac{1}{2}\log\frac{x_1}{x_2}~.
\end{equation}
The integration over $x_1$, $x_2$ can be converted to the one over $\tau$, $\rm y$ via the relations,
\begin{equation}
x_{1,2} = \sqrt{\tau}e^{\pm {\rm y}}~,~~~~~dx_1 dx_2 = d\tau d{\rm y}~.
\end{equation}
Using the identity  $\delta(\tau s - M_{ll}^2) = \delta(\tau - M_{ll}^2/s)/s$ we get rid of the integration over $\tau$, and we find
\begin{eqnarray}\label{eq:LO}
\left. \frac{1}{2\, M_{ll}}  \frac{d\sigma}{dM_{ll}}\right|_{\rm th} &=& \frac{\pi M_{ll}^2}{2s}\frac{1}{[(M_{ll}^2 - m_Z^2)^2 + \Gamma_Z^2m_Z^2]}\int_{-{\rm y}_{\rm max}}^{+{\rm y}_{\rm max}} d{\rm y}
\int_{-z_{\rm max}}^{+z_{\rm max}} dz \nonumber \\
&\times&
\sum_{q,\bar{q}}
f_q(\sqrt{\tau}e^{{\rm y}})f_{\bar{q}}(\sqrt{\tau}e^{-{\rm y}})
a^2 f_{q\bar{q}}(s_W^2, M_{ll},z)~.
\end{eqnarray}
where
$z_{\rm max} \equiv \sqrt{1-4p_{\rm T}^2/M_{ll}^2}$ and ${\rm y}_{\rm max}={\rm min}\{\log(s/M_{ll}^2)/2, {\rm y}_{\rm cut}\}$; 
the detector acceptance sets the value ${\rm y}_{\rm cut} = 2.5$. The nominal cut on the transverse momentum of the leptons is $p_{\rm T} = 25$ GeV. We use the MSTW parton distribution functions at NNLO~\cite{Martin:2009iq}. 

\item[$\circ$] In Eq.~(\ref{eq:LO}) the PDF are evaluated setting the renormalization scale at the dynamical value $\mu = M_{ll}$. Furthermore, we make use of the central PDF set.
Fixing these two values implies the introduction of both scale and PDF uncertainties in the computation of the theoretical cross-section. We estimate the impact of the scale uncertainty by varying the renormalization scale in $\mu  = [1/2\, M_{ll},\, 2 M_{ll}]$, and we find that it introduces at most $\sim 1\%$ error (see~\cite{Alves:2014cda} for the related discussion). We therefore neglect such correction in our analysis.
PDF uncertainty is larger, and can be estimated
evaluating the DY cross-section over a statistical sample obtained 
by changing the PDF eigenvector set in Eq.~(\ref{eq:LO}). We include the PDF uncertainty in our analysis, and we refer the reader to section~\ref{sec:Analysis} for a detailed 
discussion.

\item[$\circ$] Eq.~(\ref{eq:LO}) is defined at the LO in the hard scattering process.
We include NNLO corrections by properly rescaling---bin by bin in the invariant mass spectrum range---the LO 
cross-section with respect to the NNLO cross-section, relying on the numerical evaluation of Ref.~\cite{CMS:2014jea}. 

\end{itemize} 

We are now ready to compare the theoretical cross-section in Eq.~(\ref{eq:LO}) with experimental data
extracted from the measurement of the differential DY cross-section in the di-electron and di-muon channels.

\section{Analysis and Results}\label{sec:Analysis}

We extract our bounds by means of a simple $\chi^2$ analysis derived comparing theory and data.
We  start defining  the $\chi^2$ function
\begin{equation}\label{eq:Fit2}
\chi^2(\mathcal{N}, N_{eff}, M_X) =  \sum_{i,j}
\left[\mathcal{N}
\left.\frac{d\sigma}{dM_{ll}}\right|_{\rm th}  -  \left.\frac{d\sigma}{dM_{ll}}\right|_{\rm exp}
\right]_i \left( \Sigma^{-1}\right)_{ij}
\left[\mathcal{N}
\left.\frac{d\sigma}{dM_{ll}}\right|_{\rm th}  -  \left.\frac{d\sigma}{dM_{ll}}\right|_{\rm exp}
\right]_j~,
\end{equation}
where the sum runs over the invariant mass bins. The theoretical cross-section in Eq.~(\ref{eq:LO}) depends on the many free parameters, $d_X,\,N_X,\, Q_X,\, M_X$. 
To simplify the notation we further define $N_{eff} \equiv  d_X N_X Q_X^2$.
We allow for a free overall normalization $\mathcal{N}$ of the theoretical cross-section---over which we marginalize---in order to include possible unknown correlated systematic uncertainties. Denoting with $\chi^2_{\rm min}$ the minimum of the function $\chi^2(\mathcal{N},N_{eff},M_X)$, and introducing the marginalized distribution $\chi_{\rm marg}^2(N_{eff},M_X) = \int d\mathcal{N}\chi^2(\mathcal{N},\, N_{eff}, M_X)$,  we derive confidence level contours by requiring $\chi_{\rm marg}^2(N_{eff}, M_X) > \chi^2_{\rm min} + \kappa$, with $\kappa= 2.30,\,6.18,\,11.83$ at, respectively,  $1$-, $2$- and $3\,\sigma$ level.

In Eq.~(\ref{eq:Fit2}) 
$\left.d\sigma/dM_{ll}\right|_{\rm exp}$ 
represents the experimentally measured differential cross-section in the dilepton invariant mass
while
$\Sigma$ is the covariance error matrix. 
Before we proceed further,
we have to distinguish between the analysis at $\sqrt{s} = 8$ TeV and  $\sqrt{s} = 14$ TeV.
The reason is that at $\sqrt{s} = 8$ TeV we can use  data and errors from the DY analysis carried out by CMS in~\cite{CMS:2014jea}. 
At $\sqrt{s} = 14$ TeV, on the contrary, we shall rely on a projection.

\subsection{Drell-Yan at $\sqrt{s} = 8$ TeV}\label{sec:8TeV}

We use the results of the analysis presented by the CMS collaboration in~\cite{CMS:2014jea} where
the differential cross-section in the dilepton invariant mass range $M_{ll} = [15,\, 2000]$ GeV was measured 
using proton-proton collisions at $\sqrt{s} = 8$ TeV with an integrated luminosity of $19.7$ fb$^{-1}$.
Data and errors are publicly available at the website of the Durham HepData Project~\cite{hepdata_DY}.

The covariance error matrix entering in Eq.~(\ref{eq:Fit2}) is given 
by $\Sigma = \Sigma_{\rm exp} + \Sigma_{\rm PDF}$. $\Sigma_{\rm exp}$ encodes 
experimental errors (both statistical errors and systematic uncertainties, as discussed in~\cite{CMS:2014jea}) 
and correlations related to the pre-FSR invariant mass distribution in the combined dilepton 
channel~\cite{CMS:2014jea} while $\Sigma_{\rm PDF}$ takes into account 
the impact of PDF uncertainties in the computation of the differential DY cross-section. 
We take $\Sigma_{\rm exp}$ from~\cite{hepdata_DY}.
The knowledge of the covariance matrix 
is of fundamental importance since it
gives us the possibility to compute the correlation matrix describing correlations between different bins in the invariant mass distribution.
The correlation matrix can be written as
\begin{equation}\label{eq:Correlation}
\mathcal{C}_{\rm exp} = [{\rm diag}(\Sigma_{\rm exp})]^{-1/2}\Sigma_{\rm exp}\,[{\rm diag}(\Sigma_{\rm exp})]^{-1/2}~.
\end{equation}
\begin{figure}[!htb!]
\minipage{0.5\textwidth}
  \includegraphics[width=.9\linewidth]{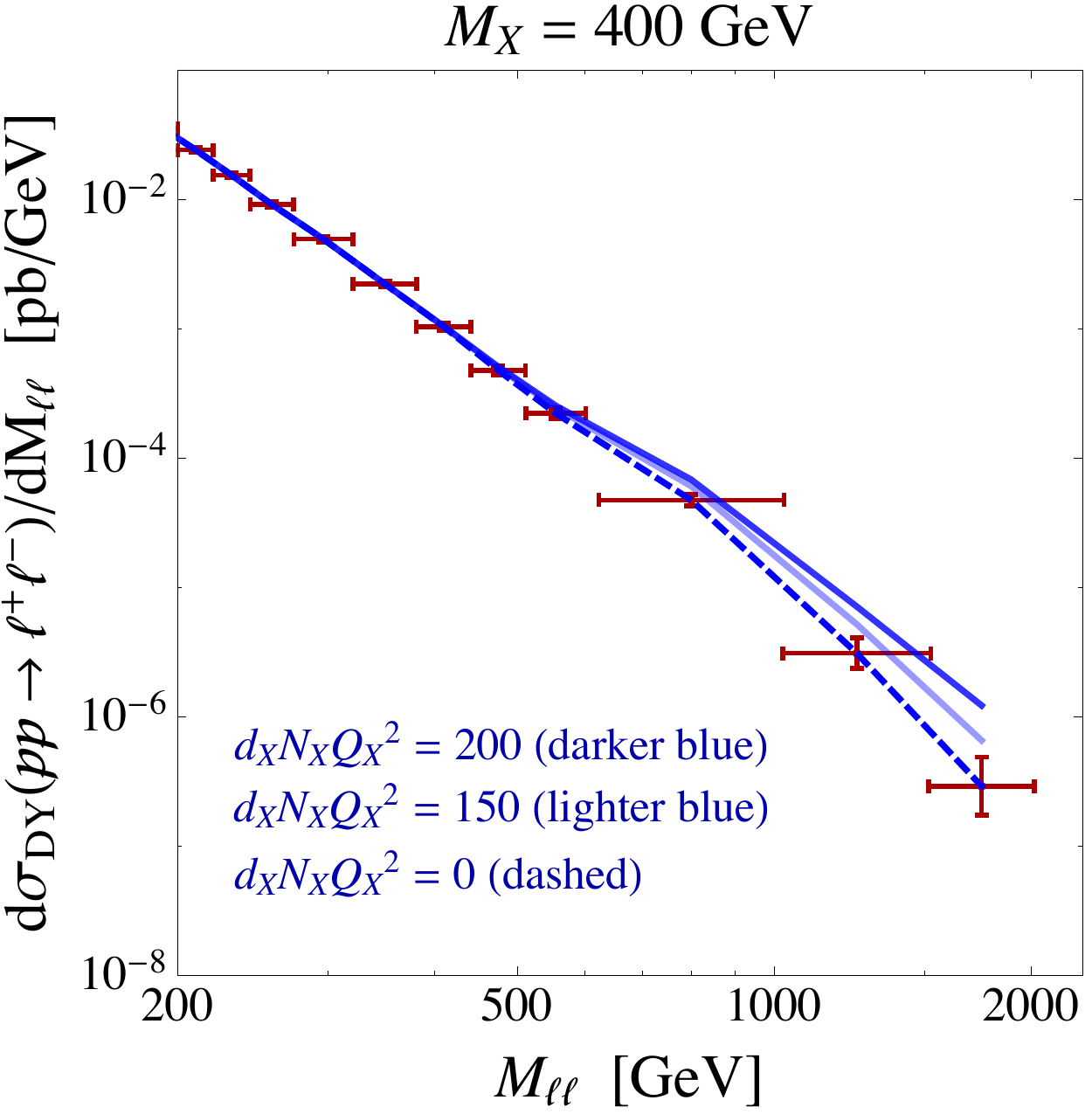}
\endminipage\hfill
\minipage{0.5\textwidth}
  \includegraphics[width=.9\linewidth]{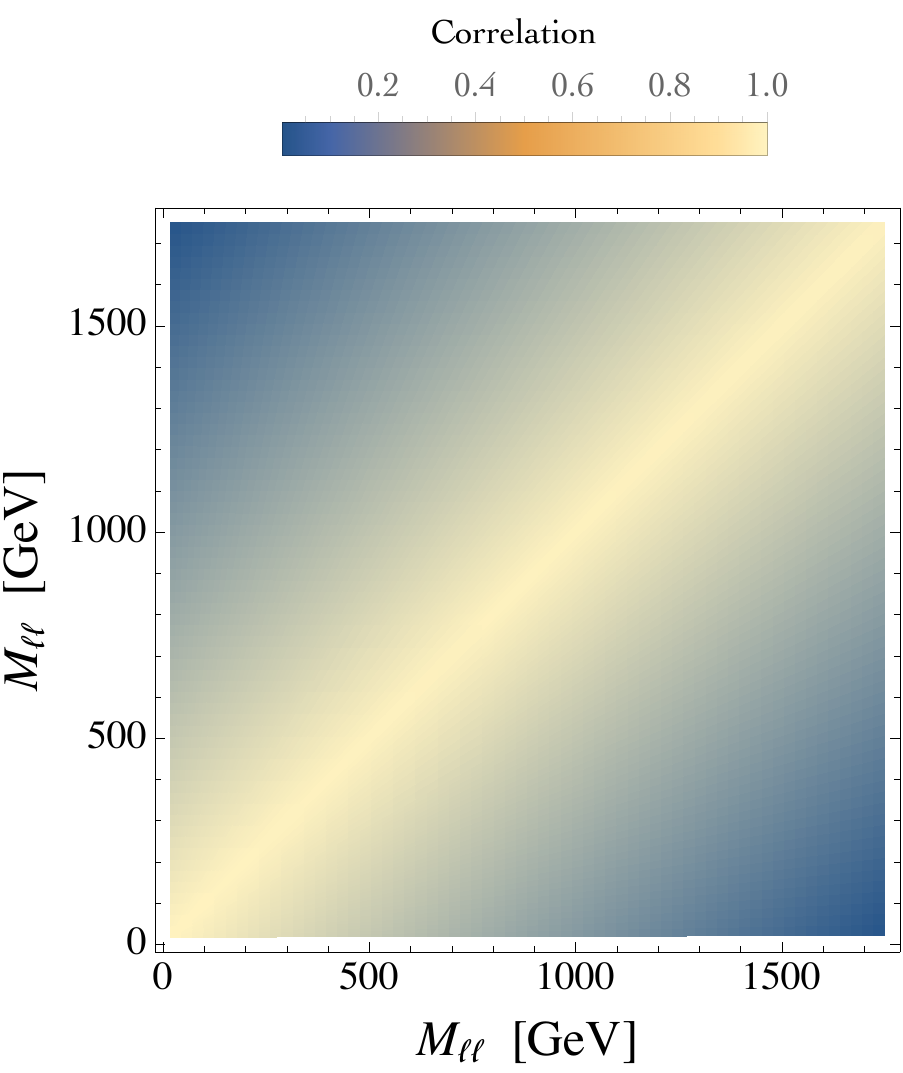}
\endminipage
\caption{\em 
Left panel. Correlation matrix in Eq.~(\ref{eq:Correlation}) derived from the experimental covariance error matrix $\Sigma_{\rm exp}$. 
Right panel. Comparison between 
data and theoretical cross-section at large dilepton invariant mass. We show the impact of running couplings 
for $M_X = 400$ GeV and different values of the combination $d_X N_X Q_X^2$. 
}
\label{fig:DYRunning}
\end{figure}
In the left panel of Fig.~\ref{fig:DYRunning} we show the comparison between experimental data 
and theoretical cross-section at large dilepton invariant mass. We simulated the SM theoretical prediction of
the DY fiducial cross section with {\tt FEWZ}~\cite{Gavin:2010az} and we explain all the details of this simulation, its
expected accuracy and comparison with the experimental CMS data in the Appendix.  
The dashed blue line corresponds to the NNLO SM cross-section, 
while the blue lines include the extra vector-like fermions.
The NNLO SM cross-section is in excellent agreement with the measured values in the whole range of dilepton invariant mass. 
Let us now discuss the impact of the new vector-like fermions.
For illustrative purposes we fix $M_X = 400$ GeV, and, to better visualize the impact of running couplings, we show
two specific cases with $d_X N_X Q_X^2 = 150$ (lighter blue) and $d_X N_X Q_X^2 = 200$ (darker blue). 
In the dilepton invariant mass range $500 \lesssim M_{ll} \lesssim 1000$ GeV the 
differential cross-section is measured with a $10\%$ accuracy.
For $M_{ll} \geqslant M_X$ the vector-like fermions actively participate to the hypercharge running, and 
their impact on the differential cross-section may easily overshoot the data points, as qualitatively shown in Fig.~\ref{fig:DYRunning}, for 
sufficiently large $d_X N_X Q_X^2$.
In the right panel of  Fig.~\ref{fig:DYRunning} we show the correlation matrix derived in Eq.~(\ref{eq:Correlation}). As expected, the plot highlights the presence of strong correlations
between adjacent bins. The inclusion of the correlation matrix in the $\chi^2$ fit plays an important role
since it allows to constraint---in addition to the absolute deviation from the observed values 
in each individual bin---also the slope of theoretical  cross-section.

Let us now discuss the size of PDF uncertainties. The differential cross-section in Eq.~(\ref{eq:LO}) was obtained considering 
the central PDF set (corresponding to the PDF best-fit). 
In order to assess the impact of PDF uncertainties we need to 
statistically quantify---using all the remaining eigenvector PDF sets---the relative change in the cross-section.
Let us discuss this point in more detail.
In order to construct the covariance error matrix $\Sigma_{\rm PDF}$
we need two ingredients
\begin{itemize}

\item[$\circ$] PDF uncertainties in individual bins;

\item[$\circ$] Correlation matrix among different bins. 

\end{itemize}
In the following we denote with $\sigma^{(k)}_{\pm}(M_{ll})$ the differential cross-section $d\sigma/dM_{ll}$ evaluated at dilepton invariant mass $M_{ll}$ using the $k^{\rm th}$ PDF eigenvector pair.
We start computing the PDF uncertainty in individual bins. We follow the standard treatment in~\cite{Martin:2009iq,Buckley:2014ana}.
The PDF uncertainty corresponds to 
\begin{equation}\label{eq:Sigma}
\mathcal{S}(M_{ll}) = \frac{1}{2}\sqrt{
\sum_{k=1}^{N}\left[
\sigma^{(k)}_{+}(M_{ll}) - \sigma^{(k)}_{-}(M_{ll})
\right]^2
}~.
\end{equation}
Correlations among different bins can be computed using standard statistics, and we find
\begin{equation}\label{eq:CorrPDF}
(\mathcal{C}_{\rm PDF})_{ij} = \frac{
\sum_{k=1}^{N}\left[
\sigma^{(k)}_{+}(i) - \sigma^{(k)}_{-}(i)
\right]\left[
\sigma^{(k)}_{+}(j) - \sigma^{(k)}_{-}(j)
\right]
}{
\sqrt{
\sum_{k^{\prime}=1}^{N}\left[
\sigma^{(k^{\prime})}_{+}(i) - \sigma^{(k^{\prime})}_{-}(i)
\right]^2
\sum_{k^{\prime\prime}=1}^{N}\left[
\sigma^{(k^{\prime\prime})}_{+}(j) - \sigma^{(k^{\prime\prime})}_{-}(j)
\right]^2
}
}~,
\end{equation}
where $i$ and $j$ denote two invariant mass bins. Equipped by these results, we can compute the covariance error matrix.
We have
\begin{equation}
\Sigma_{\rm PDF} = \mathcal{S}\times \mathcal{C}_{\rm PDF}\times \mathcal{S}~,
\end{equation}
where we defined the diagonal matrix $\mathcal{S} \equiv {\rm diag}(\mathcal{S}_i)$, with $\mathcal{S}_i \equiv \mathcal{S}(i)$.
\begin{figure}[!htb!]
\minipage{0.5\textwidth}
  \includegraphics[width=.9\linewidth]{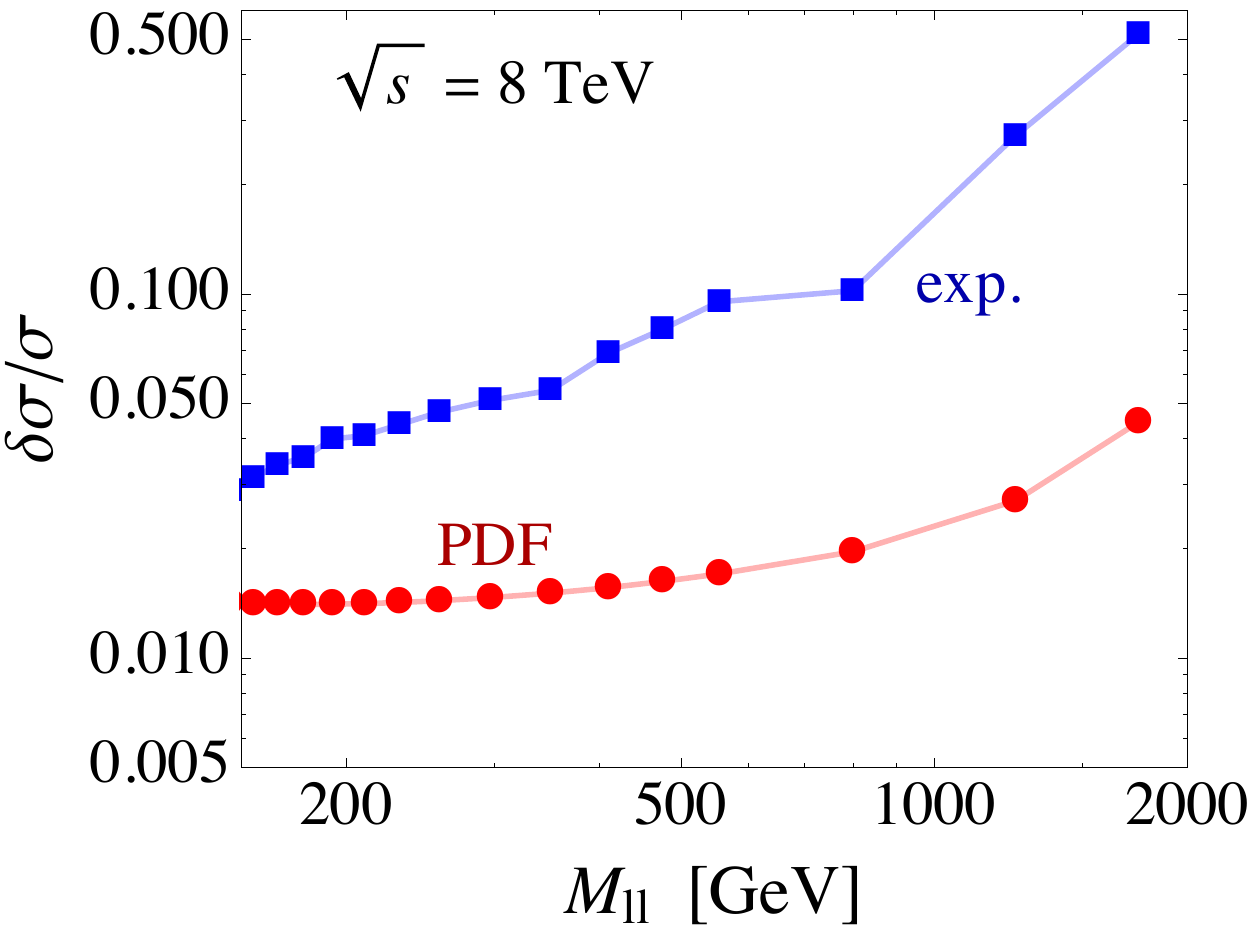}
\endminipage\hfill
\minipage{0.5\textwidth}
  \includegraphics[width=.9\linewidth]{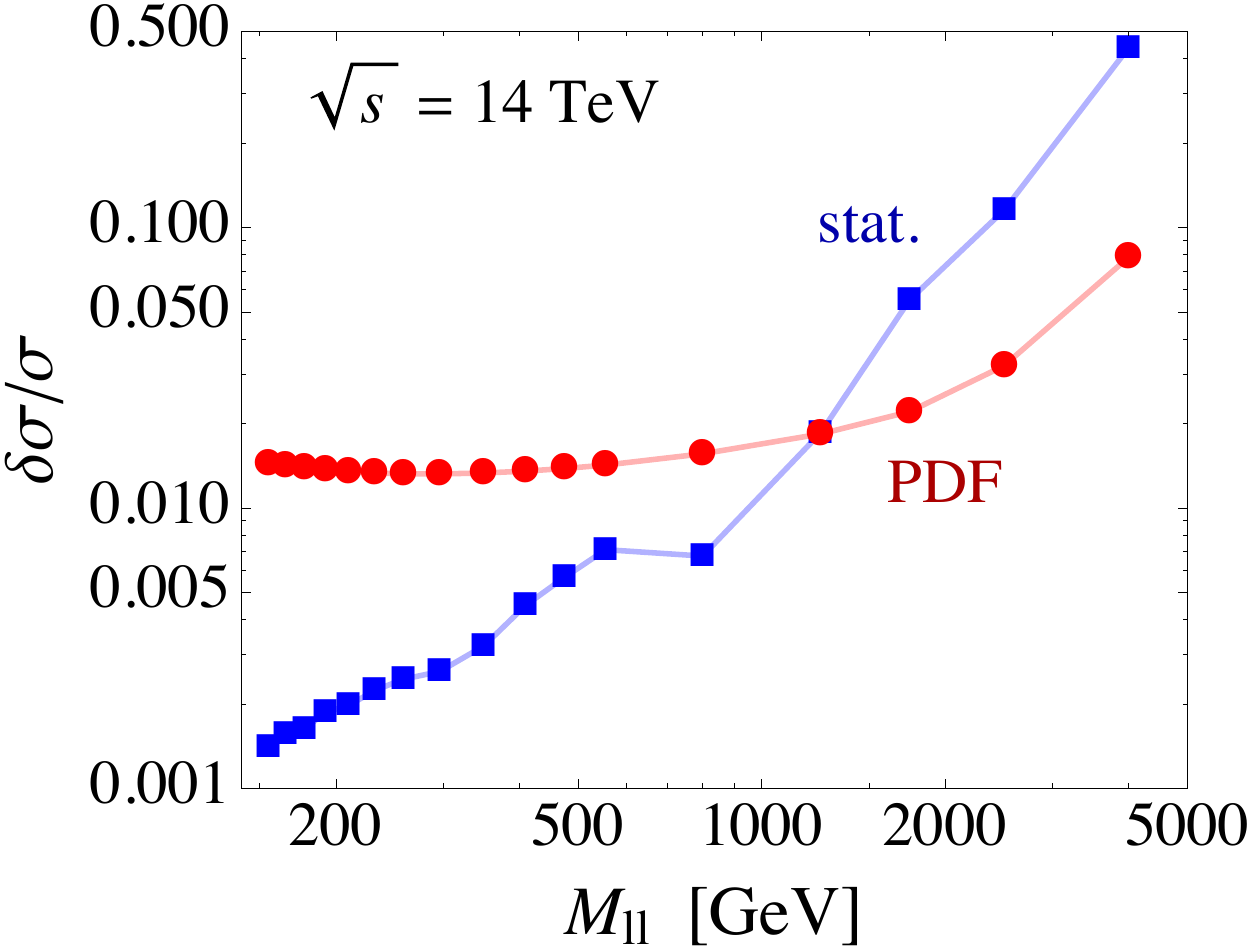}
\endminipage\\
\minipage{0.5\textwidth}
  \includegraphics[width=.85\linewidth]{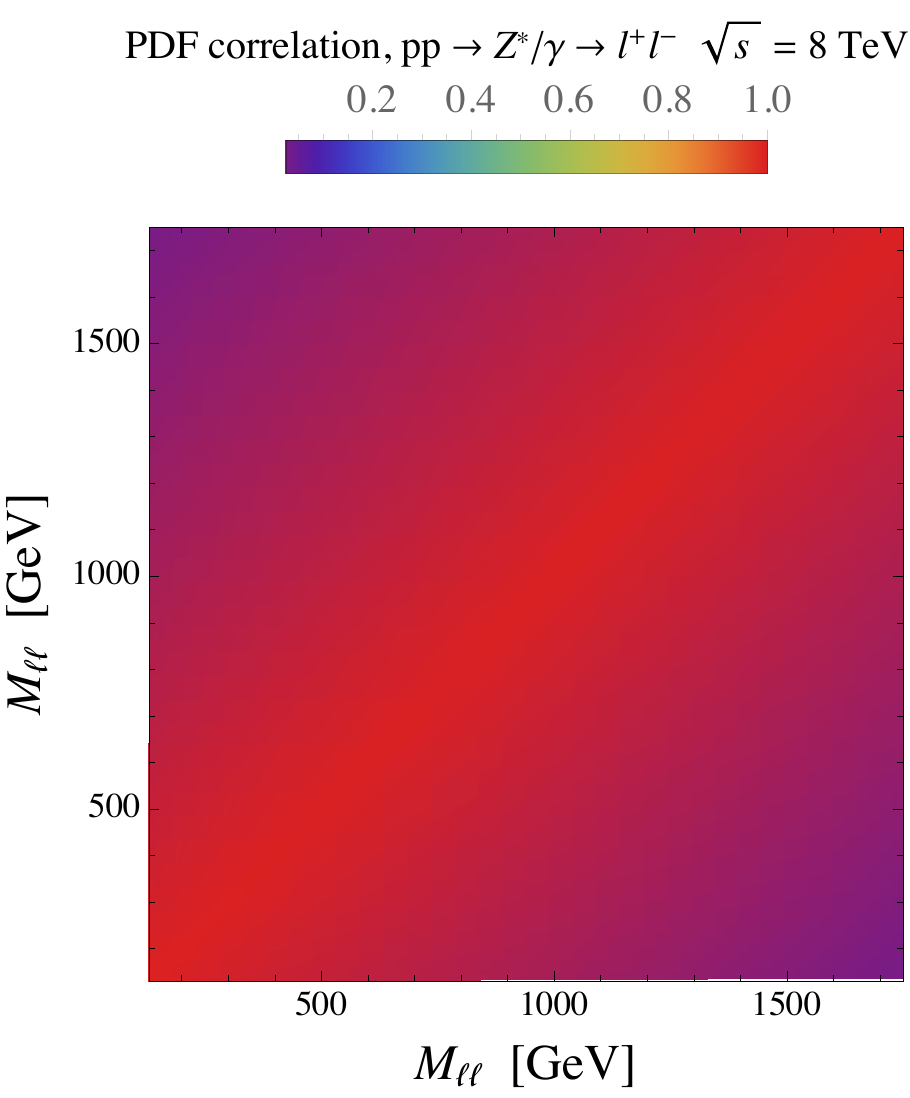}
\endminipage\hfill
\minipage{0.5\textwidth}
  \includegraphics[width=.9\linewidth]{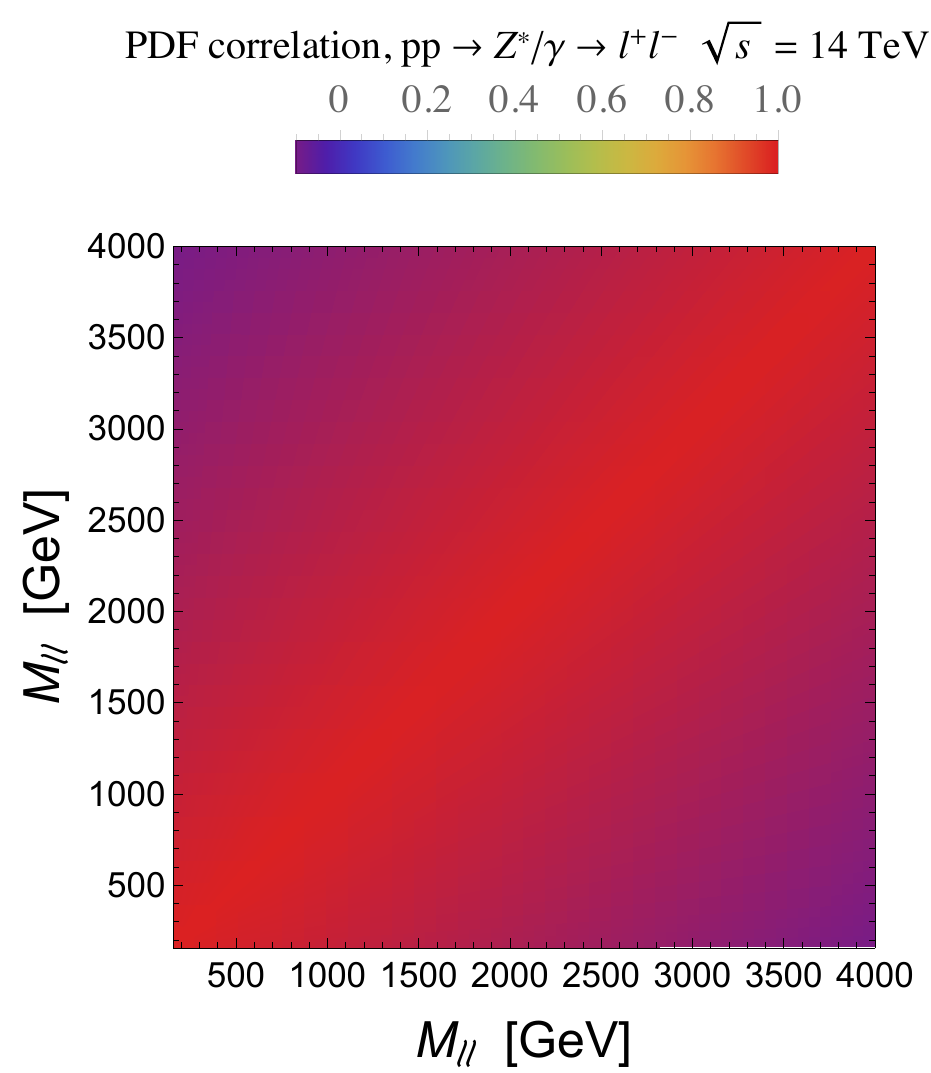}
\endminipage
\caption{\em 
Impact of PDF uncertainties. Upper row, left panel. Comparison between the experimental error (blue) and the PDF uncertainty (red) at $\sqrt{s} = 8$ TeV.
The experimental errors take into account both statistical error and systematic uncertainties, as described in~\cite{CMS:2014jea}. 
Upper row, right panel. Comparison between the statistical error (blue) and the PDF uncertainty (red) at $\sqrt{s} = 14$ TeV.
Lower row. Correlation matrix at $\sqrt{s} = 8$ TeV (left) and $\sqrt{s} = 14$ TeV (right).
}
\label{fig:PDF}
\end{figure}
We show our results in the left column of Fig.~\ref{fig:PDF}. In the upper plot, we show the relative PDF errors per individual bin (red circles), and we compare 
them with the experimental errors
quoted in~\cite{CMS:2014jea,hepdata_DY} (blue squares). As clear from this plot, at $\sqrt{s} = 8$ TeV the impact of PDF uncertainties is sub-leading if compared with the experimental errors.
For completeness, in the lower panel we show the PDF correlation extracted according to Eq.~(\ref{eq:CorrPDF}).

Let us now discuss our findings, shown in the left panel of Fig.~\ref{fig:Projection} where we plot the $1$-, $2$- and $3\, \sigma$ bound.
First of all, it is interesting to compare the DY bound with the constraint placed by LEP in Fig.~\ref{fig:EWPT}.
The net result is that the measurement of the DY differential cross-section at large invariant mass 
at the LHC with $\sqrt{s} = 8$ TeV already provides  a bound stronger than the one obtained at LEP.
This is a conceptually remarkable result, given the penalizing price unavoidably paid by an hadronic machine like the LHC 
in performing precision measurements.
Notwithstanding this important observation, it is also clear from Fig.~\ref{fig:Projection} 
that the DY bound extracted at $\sqrt{s} = 8$ TeV does not rule out any relevant portion of the parameter space since it requires at least 
$d_X N_X Q_X^2 \gtrsim 50$, a value objectively too large for any realistic model.

However, encouraged by the promising result  obtained at $\sqrt{s} = 8$ TeV, we move now to explore future prospects at 
$\sqrt{s} = 14$ TeV.

\subsection{Drell-Yan at $\sqrt{s} = 14$ TeV}

At $\sqrt{s} = 14$ TeV we have to rely on a projection.
We generate mock data assuming that the observed invariant mass distribution 
agrees with the NNLO QCD SM prediction.
We include the effect of running couplings according to Eq.~(\ref{eq:LO}),
and we construct a $\chi^2$ distribution as in Eq.~(\ref{eq:Fit2})---thus including an overall free normalization $\mathcal{N}$ as nuisance parameter to account for correlated 
unknown systematic uncertainties. 
We write the covariance error matrix as
\begin{equation}
\Sigma = \Sigma_{\rm stat} + \Sigma_{\rm uncorr\,syst} + \Sigma_{\rm PDF}~.
\end{equation}
Statistical errors are obtained assuming an integrated luminosity $\mathcal{L} = 300$ fb$^{-1}$, and converting the cross-section in terms of number of events per bin. 
The corresponding 
covariance error matrix $\Sigma_{\rm stat}$ is diagonal, with entries equal to the square of the statistical errors.
The PDF uncertainties are estimated as discussed before (see Eqs.~(\ref{eq:Sigma},\ref{eq:CorrPDF})). 
We show our results in the right column of Fig.~\ref{fig:PDF}.
Finally, 
$\Sigma_{\rm uncorr\,syst} $ is built assuming a flat $1\%$ ($2.5\%$) uncertainty across all invariant 
mass bins in order to simulate the presence of uncorrelated systematic errors.
\begin{figure}[!htb!]
\minipage{0.5\textwidth}
  \includegraphics[width=.95\linewidth]{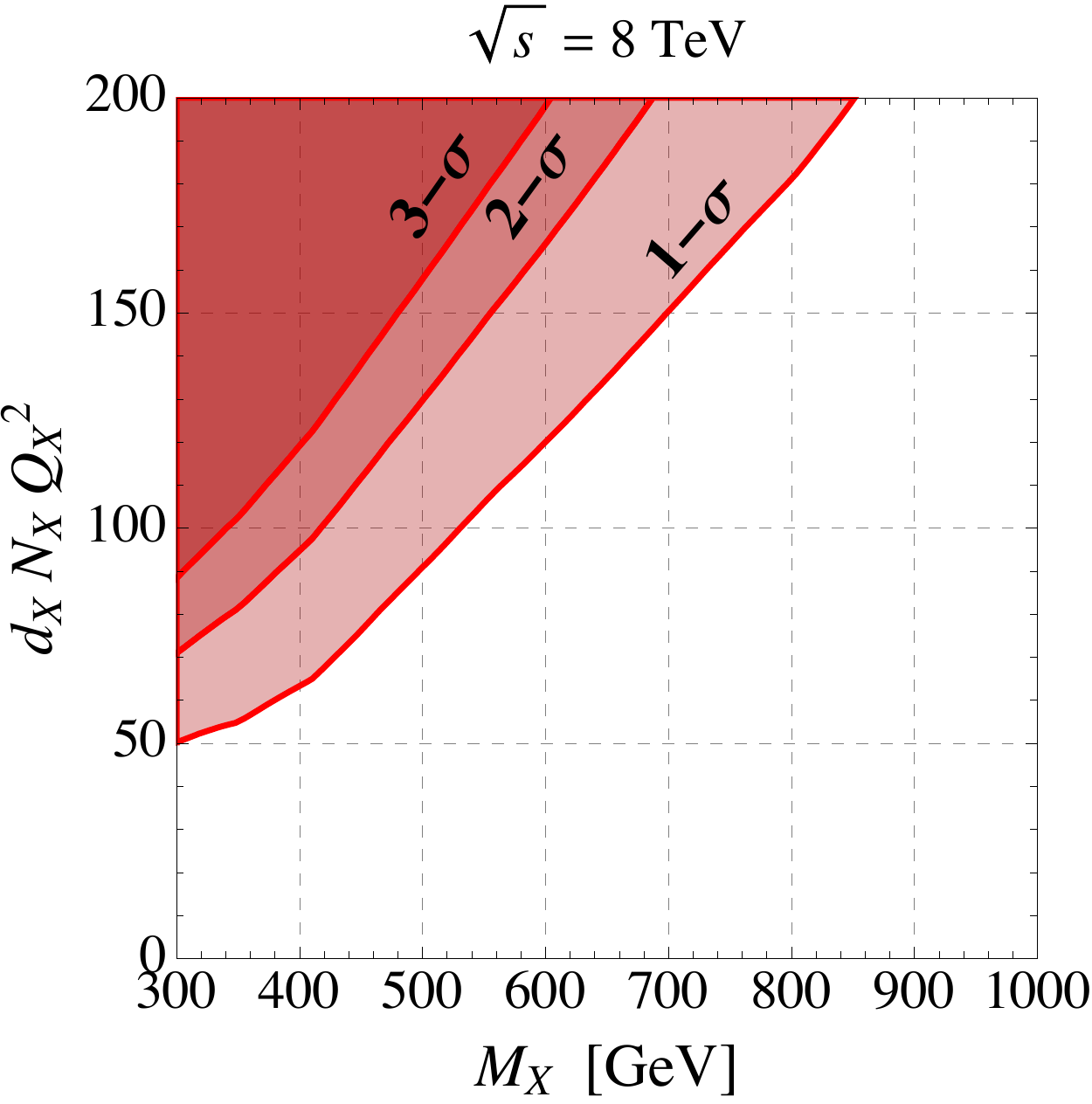}
\endminipage\hfill
\minipage{0.5\textwidth}
  \includegraphics[width=.9\linewidth]{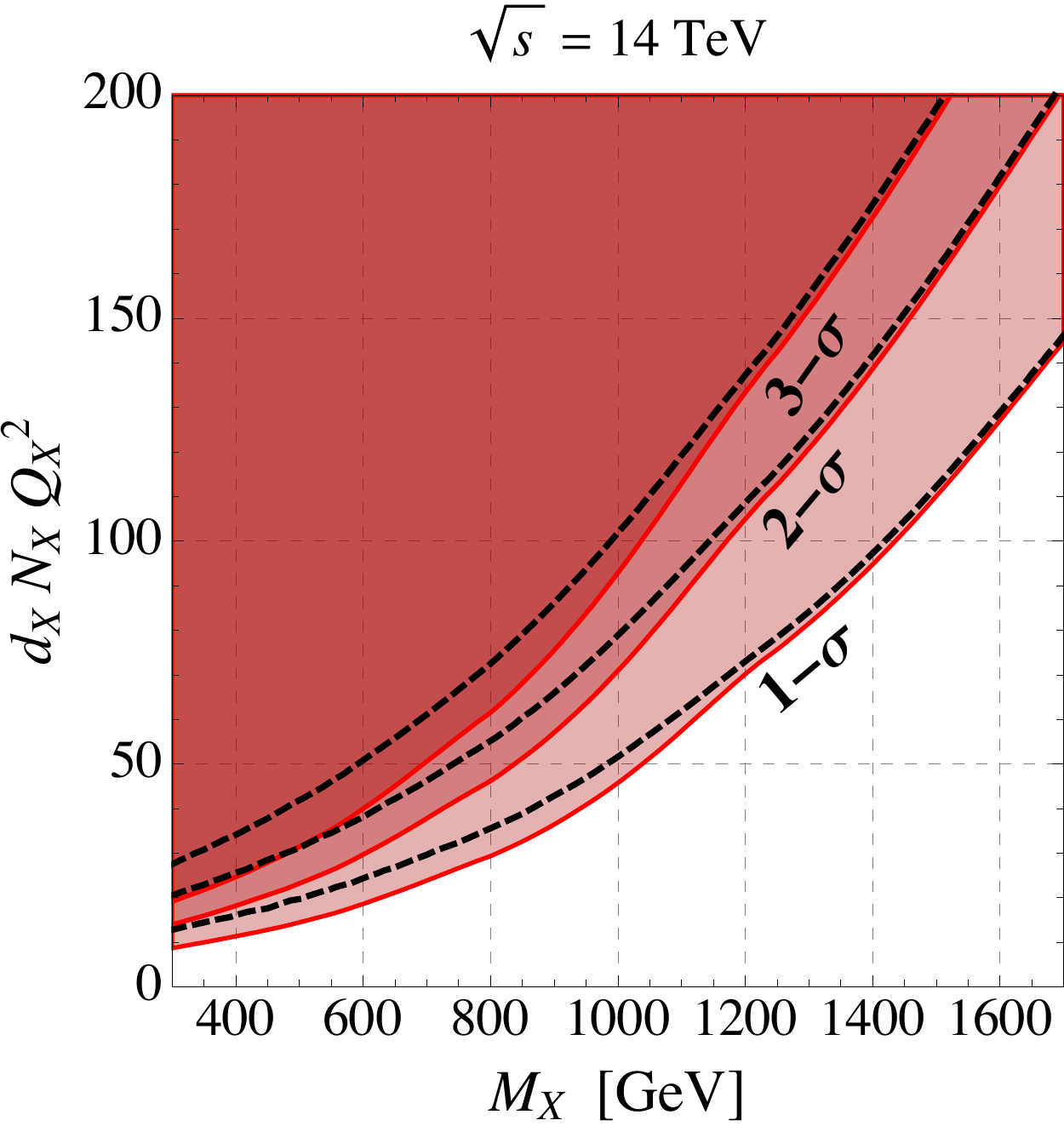}
\endminipage
\caption{\em 
Left panel.  $1$-, $2$- and $3$-$\sigma$ bound at $\sqrt{s} = 8$ TeV.
Right panel. $1$-, $2$- and $3$-$\sigma$ bound at $\sqrt{s} = 14$ TeV. Solid red (dashed black) lines were obtained including $1\%$ ($2.5\%$) uncorrelated systematic error. 
}
\label{fig:Projection}
\end{figure}
At $\sqrt{s} = 14$ TeV the simulated data extend up to $M_{ll} \simeq 5$ TeV; the statistical error, assuming $\mathcal{L} = 300$ fb$^{-1}$, 
does not exceed the $10\%$ level up to $M_{ll} \simeq 2$ TeV as shown in the upper-right panel of Fig.~\ref{fig:PDF}.
At invariant mass $M_{ll} \lesssim 1$ TeV the statistical error stays below $1\%$, and the PDF error, as well as the uncorrelated systematic error,  starts to become important.
From this simple estimate  it is clear that the constraining power of the 
DY differential cross-section at $\sqrt{s} = 14$ TeV will lead to much stronger bounds w.r.t. those obtained in section~\ref{sec:8TeV}
using data 
 at
 $\sqrt{s} = 8$ TeV.
 
 We show our findings in the right panel of Fig.~\ref{fig:Projection} where we plot the $1$-, $2$- and $3\, \sigma$ bound.
 The solid (dashed) contours in red (black) were obtained considering $1\%$ ($2.5\%$) uncorrelated systematic uncertainties.
 The impact of these uncorrelated systematic uncertainties is particularly important for light vector-like fermions and small values of $d_X N_X Q_X^2$, where 
  the deviation due to running couplings is smaller and thus it can be more easily hidden in the uncertainty accompanying the measured cross-section. 
  Despite this, it is clear that the constraining power (or, said differently, the discovery potential) of the DY process at the LHC with $\sqrt{s} = 14$ TeV and $300$ fb$^{-1}$
  starts to bite into an  interesting region of the parameter space 
  where, in particular for light vector-like fermions, the effective coupling  $d_X N_XQ_X^2$ 
  is lowered down to phenomenologically realistic values. 
  In section~\ref{sec:Future} we will discuss the implication of the DY process for the diphoton excess at $750$ GeV.
 
 It is possible to speculate even further about the role of the DY process as precision observable at the LHC.
As clear from the upper right panel of Fig.~\ref{fig:PDF}, the dominant error at large dilepton invariant mass comes from limited statistics.
The High Luminosity LHC (HL-LHC) program has the scope of attaining the threshold of $3000$ fb$^{-1}$ of integrated luminosity. 
In this case, reducing the statistical error by roughly a factor $10$, it would be possible to largely improve the constraining power/discovery potential of the DY channel even for vector-like states with TeV-scale mass.

\subsection{New charged scalars}
\label{sec:scalars}

Before proceeding, let us briefly discuss the case in which the SM is extended by adding $N_X$ complex singlet scalars $X$ with mass $M_X$ and 
hypercharge  $Y = Q_X$.
For a more general discussion we refer the reader to~\cite{Salvio:2016hnf}. At one loop, the contributions to the $\mathcal{Y}$ parameter and the hypercharge one-loop $\beta$ function are
\begin{equation}\label{eq:ScalarCase}
\mathcal{Y} = d_X N_X Q_X^2 \, \frac{\alpha_Y m_W^2}{120\pi M_X^2}~,~~~~
\beta_{g_1}^{(1)} = \left(
\frac{41}{10} + d_X N_X Q_X^2 \frac{1}{5}
\right) g_1^3~,
\end{equation}
where, in parallel with the fermionic case, the number of effective degrees of freedom $d_X N_{X}$ accounts for possible color multiplicity.
If compared to the fermionic case, the scalar contribution to the $\mathcal{Y}$ parameter turns out to be eight times smaller, and 
the bound from LEP becomes irrelevant even for very light scalar particles. 
In the absence of any constraint from LEP, it is important to assess the constraining power of the DY analysis at the LHC.
Using the fermionic case as basis for comparison, from Eq.~(\ref{eq:ScalarCase}) we see that the impact of a charged scalar 
particle on the $g_1$ running is four times smaller. We therefore expect a weaker bound from the DY analysis.
To fix ideas, at $\sqrt{s} = 8$ TeV and for $M_X = 400$ GeV
 we find that the running induced by new charged scalar particles produces  a 
 $10\%$ correction to the differential cross-section at $M_{ll} = 800$ GeV only for $d_X N_X Q_X^2 \simeq 270$.
 Contrarily, at $\sqrt{s} = 14$ TeV and for $M_X = 400$ GeV 
 a $2\%$ deviation at $M_{ll} = 1000$ GeV can be obtained with $d_X N_X Q_X^2 \simeq 50$.
 We therefore conclude that the constraining power of the DY process in the scalar case is only marginally relevant even considering 
 collisions at $\sqrt{s} = 14$ TeV with large integrated luminosity.
 
 \subsection{FCC-ee}

The FCC-ee
 is a high-luminosity, high-precision $e^+e^-$ 
 circular collider envisioned in a new $80-100$ km tunnel in the Geneva area
 with a centre-of-mass energy from $90$ to $400$ GeV~\cite{TLEP}.
Thanks to its clean experimental environment, the FCC-ee collider
could explore the EW physics with unprecedented accuracy, 
allowing for a careful scrutiny of new physics models predicting new particles at the TeV scale and beyond.
In the following, we base our discussion on~\cite{Gomez-Ceballos:2013zzn}. 
\begin{figure}[!htb!]
\minipage{0.5\textwidth}
  \includegraphics[width=.85\linewidth]{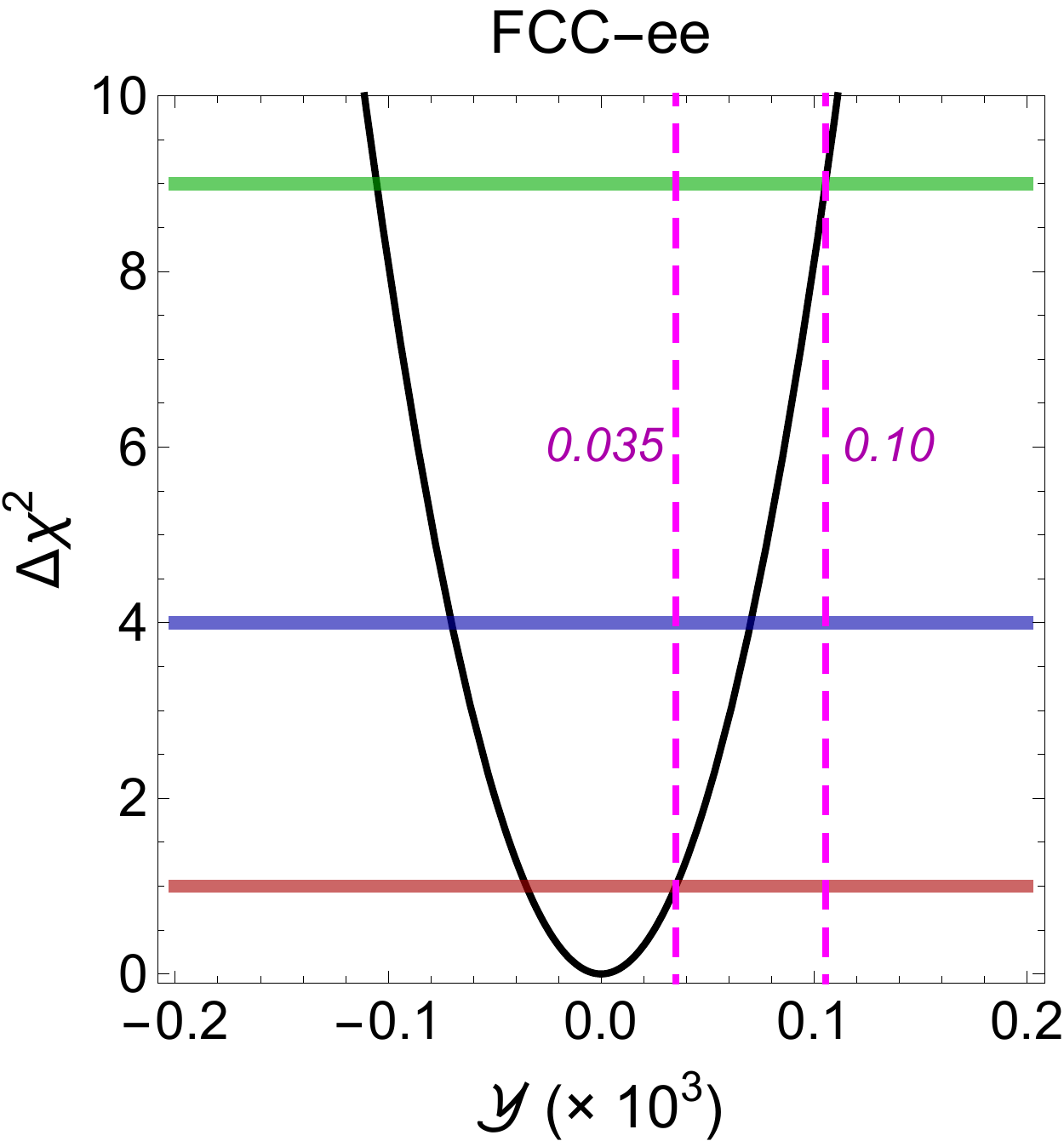}
\endminipage\hfill
\minipage{0.5\textwidth}
  \includegraphics[width=.95\linewidth]{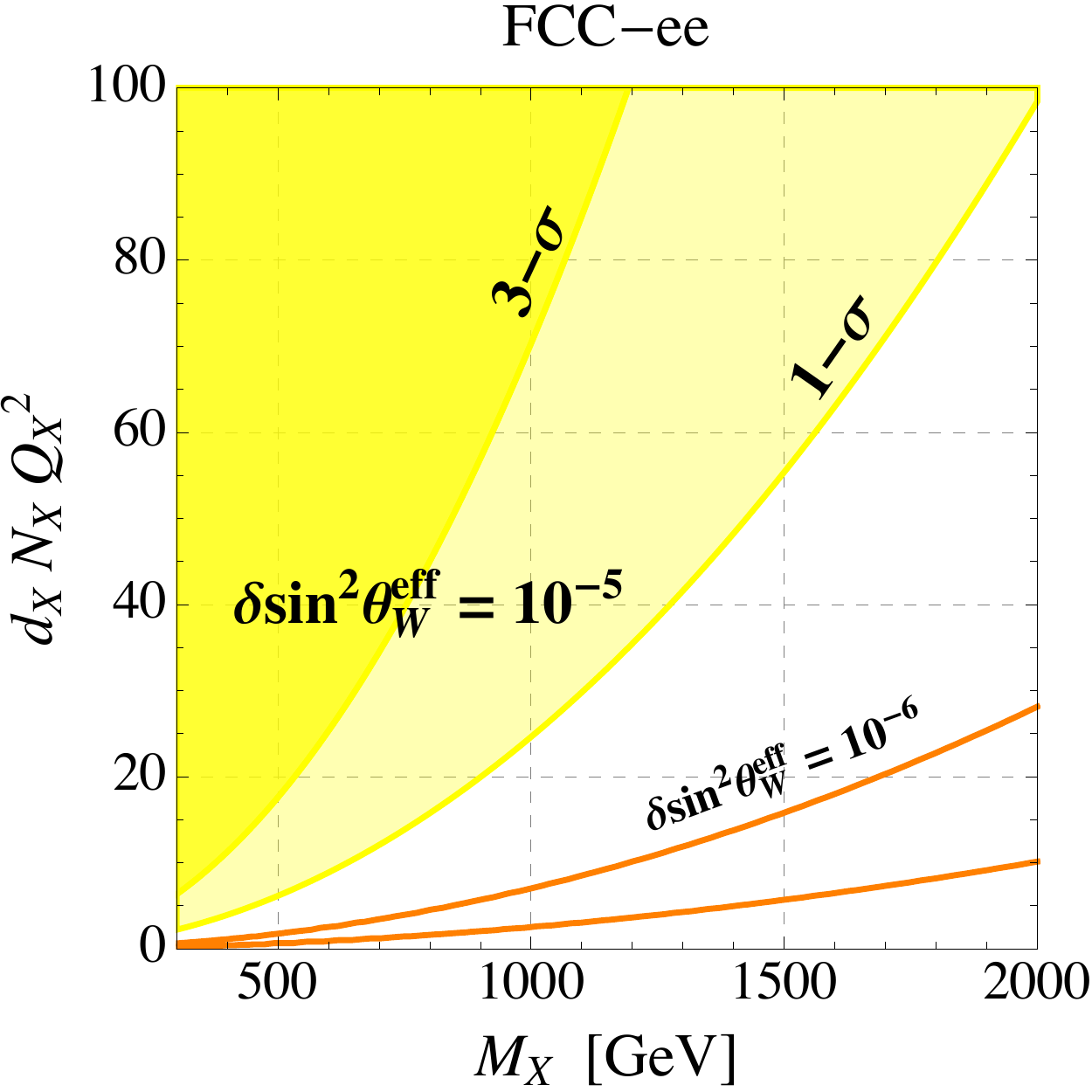}
\endminipage
\caption{\em 
Left panel. One-parameter fit of LEP data in terms of the $\mathcal{Y}$ parameter. We rescale the errors of the EW precision measurements 
according to what expected in~\cite{Gomez-Ceballos:2013zzn} for a future circular $e^+e^-$ collider in order to provide a flavor of its constraining power (see text for details). 
We set the precision on the effective squared mixing angle to $\delta\sin^2\theta_W^{\rm eff} = 10^{-5}$.
The central value of  $\mathcal{Y}$ is arbitrarily set to zero.
Right panel. $1$- and $3$-sigma exclusion regions in terms of the parameter space $(d_X N_XQ_X^2, M_X)$ (yellow).
We also show the corresponding bounds if $\delta\sin^2\theta_W^{\rm eff} = 10^{-6}$ (orange).
}
\label{fig:TLEP}
\end{figure}
To give an idea of the constraining power of the FCC-ee collider, it is possible to recast the LEP results.
For instance, according to the estimates presented in~\cite{Gomez-Ceballos:2013zzn}, at the FCC-ee it will be possible to measure 
the pole mass of the Z-boson with a precision twenty times smaller than the one reached at LEP, $\left.\delta m_Z\right|_{\rm LEP} = 0.0021$ GeV.
One can fit the LEP data implementing all the upgraded precisions quoted in~\cite{Gomez-Ceballos:2013zzn}. Following this logic, 
we show in the left panel of Fig.~\ref{fig:TLEP} the bound obtained considering a one-parameter fit made in terms of the $\mathcal{Y}$ parameter.
The most important measurement controlling the precision on $\mathcal{Y}$ 
is the square of the effective weak mixing angle. Assuming $\delta\sin^2\theta_W^{\rm eff} = 10^{-5}$, we find $\delta\mathcal{Y} = 0.035\times 10^{-3}\,(0.1\times 10^{-3})$
at $1$-$\sigma$ ($3$-$\sigma$). In the right panel of Fig.~\ref{fig:TLEP} we translate these confidence 
regions in the parameter space $(d_X N_X Q_X^2, M_X)$ (regions shaded in yellow).
As expected the constraining power of the FCC-ee collider turns out to be much stronger than the capabilities of present and future LHC searches in the DY channel.
If  $\delta\sin^2\theta_W^{\rm eff} = 10^{-6}$ (the value envisaged in~\cite{Gomez-Ceballos:2013zzn}) the constraint becomes even stronger, and we find 
 $\delta\mathcal{Y} = 0.0036\times 10^{-3}\,(0.01\times 10^{-3})$ (orange lines in the right panel of Fig.~\ref{fig:TLEP}).

\section{Implications for the Diphoton Excess at $750$ GeV}\label{sec:Future}

We are now in the position to comment about the importance of the DY process at the LHC for the diphoton excess discussed in section~\ref{sec:EWPT}.
The Yukawa interactions between the (pseudo-) scalar resonance $S$ and the vector-like fermions $X$ are encoded in the Lagrangian 
\begin{equation}
\mathcal{L} = \mathcal{L}_{SM} + \frac{(\partial_{\mu}S)^2}{2} + \bar{X}(i\slashed{D} - M_X)X - \left[
S\bar{X}(y_X + i\tilde{y}_X\gamma^5)X + h.c.
\right] - V(S) - V(S,H)~,
\end{equation}
where $\mathcal{L}_{SM}$ is the SM Lagrangian and
where we assumed that the $N_X$ fermions have the same mass and couplings.
The Yukawa coupling $y$ ($\tilde{y}$) is present 
if $S$ is a scalar (pseudo-scalar). The potential is $V(S) = (M_S^2/2) S^2 + \lambda_S S^4$, and $V(S, H)$ accounts for possible interactions with the Higgs doublet $H$. 
The explicit form of $V(S, H)$  is not important for the purposes of our discussion, and we refer to~\cite{Salvio:2016hnf} for a more general analysis including vacuum stability.
\begin{figure}[!htb!]
\minipage{0.5\textwidth}
  \includegraphics[width=.9\linewidth]{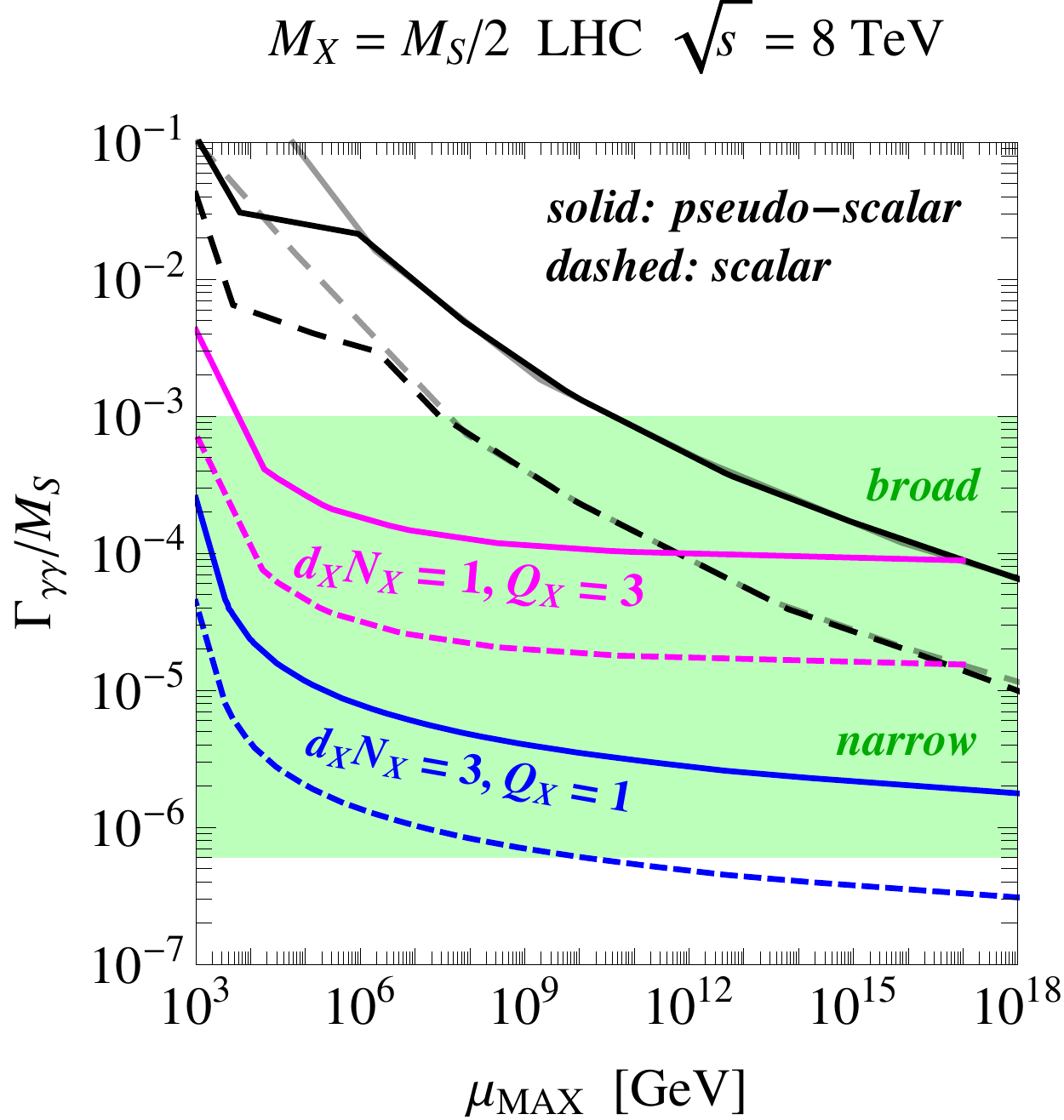}
\endminipage\hfill
\minipage{0.5\textwidth}
  \includegraphics[width=.9\linewidth]{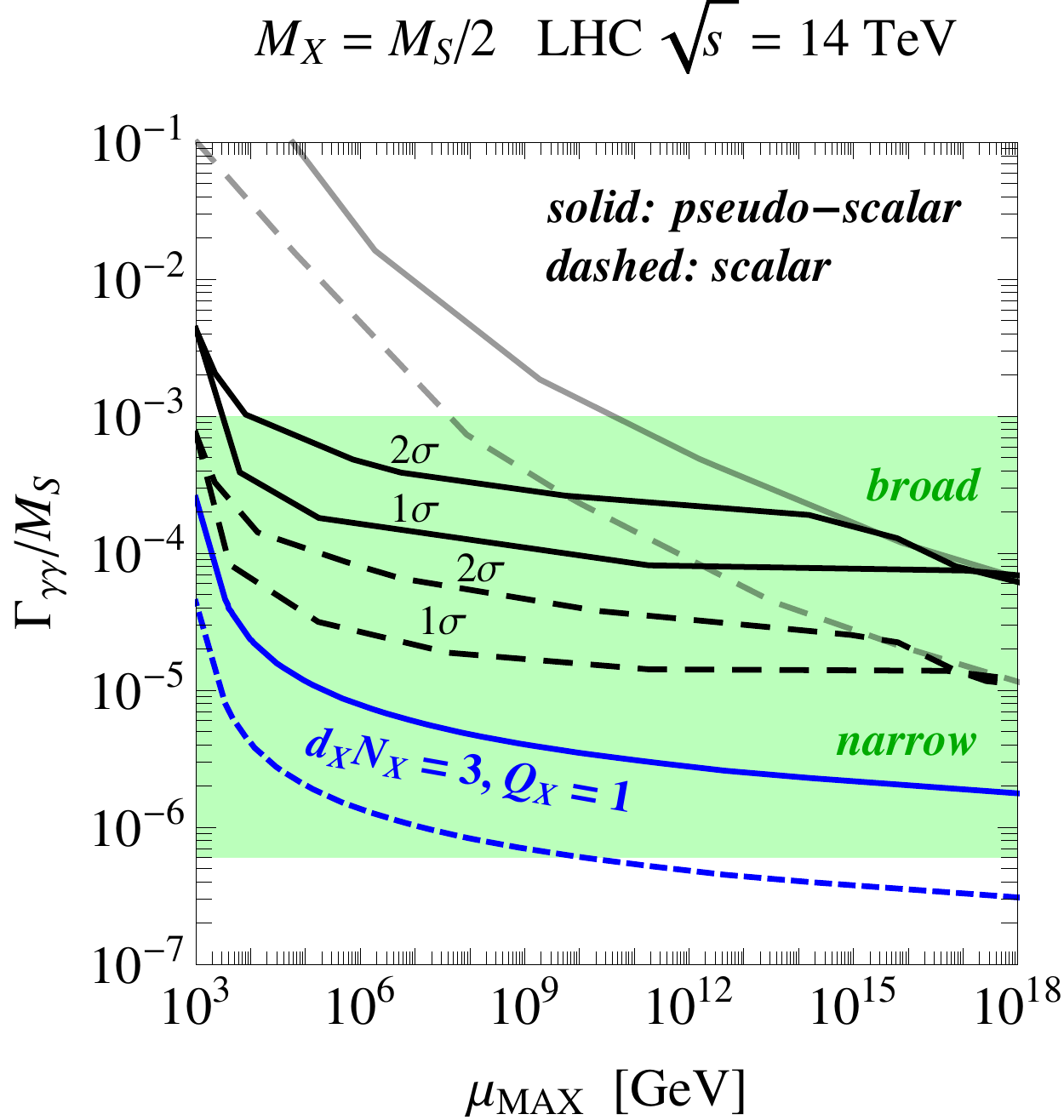}
\endminipage
\caption{\em 
Maximal diphoton decay width for a pseudo-scalar (solid lines) and scalar (dashed lines) resonance as a function 
of the scale at which the theory becomes non-perturbative.
The gray lines include the LEP bound on the $\mathcal{Y}$ parameter at $1$-$\sigma$ level.
The black lines include the LHC bound from DY at $\sqrt{s} = 8$ TeV ($1$-$\sigma$ contour,  left panel) $\sqrt{s} = 14$ TeV ($1$- and $2$-$\sigma$ contours, right panel).
We also show two particular cases with $d_X N_X = 3$, $Q_X = 1$ (blue), $d_X N_X = 1$, $Q_X = 3$ (magenta). 
}
\label{fig:GammaGammaExcess}
\end{figure}
As pointed out in~\cite{Gu:2015lxj, Goertz:2015nkp, Son:2015vfl}, it is important to keep track of the RG evolution of the Yukawa coupling since it can easily 
exceed the perturbative regime. The Yukawa couplings $y_X$ and $\tilde{y}_X$ have the same RGE, and we find\footnote{For simplicity, we do not include the RGE for $\lambda_S$ since it does not change qualitatively  our conclusions. The role
 of $\lambda_S$ related to the stability of the EW vacuum was discussed in~\cite{Son:2015vfl,Salvio:2016hnf}.}
\begin{equation}
\beta_{y_X}^{(1)} = \left[
\left(
2d_X N_X + 3\right)y_X^2 - \frac{18}{5}Q_X^2 g_1^2
\right]y_X~.
\end{equation}
The CP-odd coupling $\tilde{y}_X$ contributes to the diphoton decay width of $S$ more
than the CP-even coupling $y_X$. For the latter, we have
\begin{equation}\label{eq:Diphoton}
\frac{\Gamma_{\gamma\gamma}}{M_S} = \frac{\alpha^2}{16\pi^3}\left|
d_X N_X Q_X^2 y_X \sqrt{\tau_X}  \mathcal{S}(\tau_X)
\right|^2~,
\end{equation}
with $\tau_X \equiv 4M_X^2/M_S^2$. The pseudo-scalar case corresponds to the substitutions $y_X \to \tilde{y}_X$, $\mathcal{S}(\tau_X) \to 
\mathcal{P}(\tau_X)$.

In Fig.~\ref{fig:GammaGammaExcess} we show the maximal diphoton decay width of the (pseudo-) scalar resonance $S$
as a function of the scale at which the theory becomes non-perturbative.
In more detail, our logic goes as follows.
{\it i)} For a given value of $d_X N_X$, $Q_X$ and $y_X$ ($\tilde{y}_X$) we compute the scale at which the  theory becomes non-perturbative
by solving the RGEs for $g_1$ and $y_X$ ($\tilde{y}_X$). The x-axes in Fig.~\ref{fig:GammaGammaExcess} is therefore formally 
defined 
as the scale $\mu_{\rm max}$ at which either $y_X (\tilde{y}_X) = 4\pi$ (perturbativity of the Yukawa coupling) or $g_1 = \infty$ (hypercharge Landau pole)
 is realized along the RG flow. {\it ii)} By scanning over $d_X N_X$, $Q_X$ and $y_X$ ($\tilde{y}_X$), and using Eq.~(\ref{eq:Diphoton}),
 we can compute the maximal diphoton decay width that the (pseudo-) scalar resonance $S$ can obtain 
 for a given $\mu_{\rm max}$.
 To give an even more clear understanding, we also show in
  Fig.~\ref{fig:GammaGammaExcess} the maximal $\Gamma_{\gamma\gamma}$ corresponding to fixed values of $d_XN_X$, $Q_X$. 
 For a fixed value of $d_XN_X$, $Q_X$ one has the freedom to move on the corresponding line by changing the Yukawa coupling.
Large Yukawa couplings correspond to the left-most part of the plot, where the diphoton width $\Gamma_{\gamma\gamma}$ rapidly 
increases (being proportional to 
$y_X^2$ ($\tilde{y}_X^2$)) at the prize to lower the cut-off scale of the theory down to the TeV range.
 {\it iii)} Finally, we impose---at each point $d_X N_X$, $Q_X$---the LEP and LHC constraint derived in section~\ref{sec:EWPT} and section~\ref{sec:Analysis}.
 In the left (right) panel of Fig.~\ref{fig:GammaGammaExcess} we include the impact of the $1$- and $2$-$\sigma$ bounds derived from the analysis 
 at $\sqrt{s} = 8$ TeV ($\sqrt{s} = 14$ TeV): Every point in the scan violating such bound is discarded.
 
 Following the logic explained above, 
 in Fig.~\ref{fig:GammaGammaExcess} we show the maximal diphoton decay width for a scalar (dashed lines) and pseudo-scalar (solid lines) resonance $S$
 as a function of the cut-off scale $\mu_{\rm max}$.  
 For simplicity, we fix $M_X = M_S/2$ since this value maximizes the diphoton width.
 Gray lines include only the bound from LEP, while the black lines include the bound extracted by the DY analysis at the LHC. 
 The cases with $d_XN_X = 3$, $Q_X = 1$ and $d_XN_X = 1$, $Q_X = 3$ are shown in blue and magenta.
We are interested in values $10^{-6} \lesssim \Gamma_{\gamma\gamma}/M_S\lesssim 10^{-3}$, 
where the left (right) part of the disequality corresponds to narrow (large) width, as discussed in section~\ref{sec:EWPT}.
In this respect, the bound from LEP plays non role in constraining phenomenologically interesting values of $\Gamma_{\gamma\gamma}/M_S$.
At $\sqrt{s} = 8$ TeV, the DY bound bites into a small corner of the parameter space, as clear from the comparison 
between the gray and the black lines in the left panel of Fig.~\ref{fig:GammaGammaExcess}.
However, it does not have any relevant implications w.r.t. the diphoton excess.  
At $\sqrt{s} = 14$ TeV, the story changes.
The DY bound starts to become relevant.
In the scalar case the black line is lowered  down to $ \Gamma_{\gamma\gamma}/M_S \simeq 10^{-4}-10^{-5}$ (at, respectively, $2$- and $1$-$\sigma$), 
thus affecting the whole region 
of the parameter space favored by the large width assumption.
The pseudo-scalar case gives a similar result, with the maximal diphoton width lowered down to $ \Gamma_{\gamma\gamma}/M_S \simeq 10^{-3}-10^{-4}$.

To conclude, 
we argue that the new physics involved in the explanation of the diphoton excess at $750$ GeV
could leave---especially if the indications in favor of a large width will be confirmed---a footprint in 
the differential cross-section of the neutral 
DY process at large dilepton invariant mass.

\section{Comments on direct searches for the new states}
\label{sec:DS}
In this section let us shortly comment on the direct detection of the new states. Of course, direct detection reach 
strongly depends on the particular decay modes of the vector like fermions.
 Comparing these 
searches to the indirect searches via DY production is by no mean straightforward. Surveying 
all various possibilities
in terms of hypercharge and decay chains  is well beyond the scope of this 
paper. Therefore, we will just emphasize several points, which are generic to the EW production. 
By drawing analogy to the existing searches for the EW pair-produced states we will merely try 
to give a flavor of  what might the ballpark of the direct detection bounds.

\begin{figure}[!htb!]
\minipage{0.5\textwidth}
  \includegraphics[width=.9\linewidth]{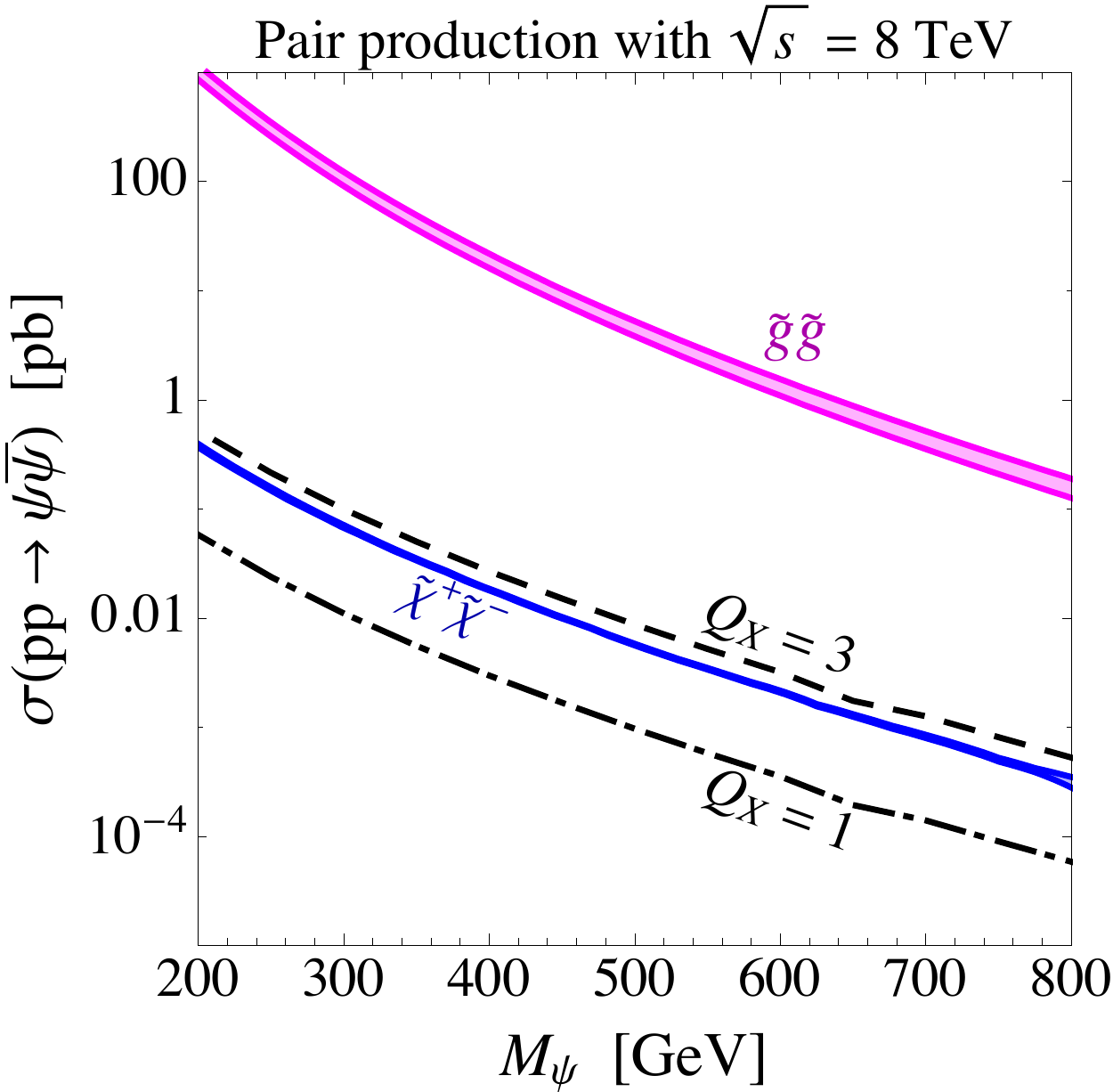}
\endminipage\hfill
\minipage{0.5\textwidth}
  \includegraphics[width=.9\linewidth]{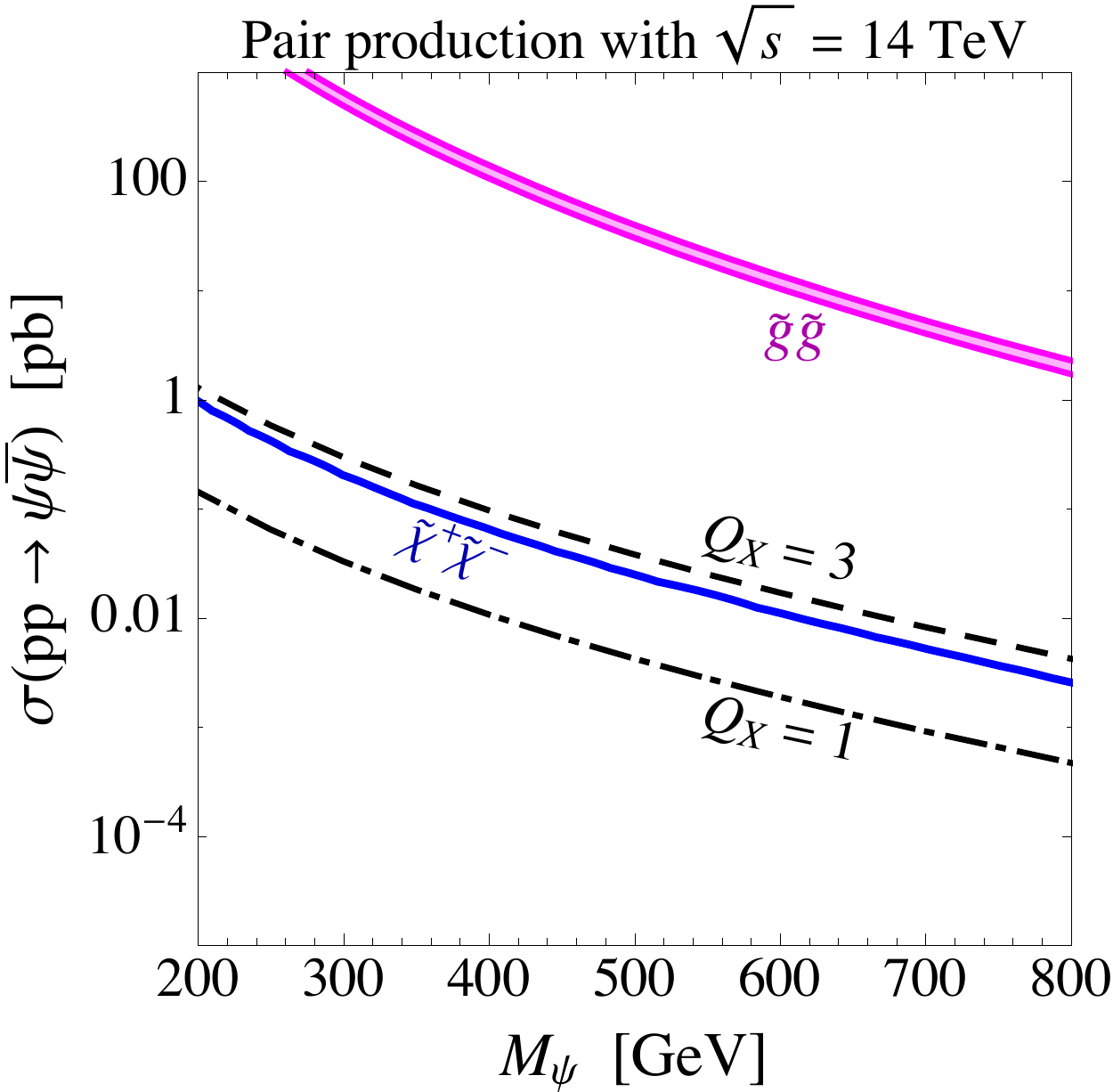}
\endminipage
\caption{\em 
Cross-section $pp\to X\bar{X}$ for producing two fermions with charge $Q_X$ at $\sqrt{s} = 8$ TeV (left panel) 
and $\sqrt{s} = 14$ TeV (right panel). The blue (magenta) 
line stands for the production cross sections of wino charginos (gluinos) in SUSY. 
}
\label{fig:Direct}
\end{figure}

At the partonic level, the cross section for the on-shell production of a $X\bar{X}$ pair is 
\begin{equation}
\sigma_{q\bar{q}}(\hat{s}) = \frac{\pi \alpha^2 Q_X^2}{162 c_W^2(s-M_Z^2)^2}
\sqrt{1-\frac{4M_X^2}{\hat{s}}}
s\left(
1+\frac{2 M_X^2}{s}
\right)\mathcal{P}_{q\bar{q}}\left(\frac{M_Z^2}{s}\right)~,
\end{equation}
where
\begin{eqnarray}
\mathcal{P}_{u\bar{u}}(x) &\equiv&  17 - 40c_W^2 x + 32 c_W^4 x^2~,\\
\mathcal{P}_{d\bar{d}}(x) &\equiv& 5 -4c_W^2 x + 8c_W^4 x^2~,
\end{eqnarray}
for, respectively, up- and down-type quarks in the initial partonic state.
The cross-section $pp \to X\bar{X}$ for producing a vector-like fermion pair with charge $Q_X$ at the LHC with $\sqrt{s} = 8,\,14$ TeV 
is shown in Fig.~\ref{fig:Direct}.

Naively one would expect that the production cross sections are EW, not that different, for example, from SUSY EW-ino
production cross sections. In general it is right, however the cross sections that we get \emph{for a single state} are generally smaller than typical EW cross sections.
The pair-production cross section of a single, charge-one state, is smaller than the 
production cross sections of the SUSY charged winos ({\rm cf. } Fig.~\ref{fig:Direct}). The explanation is very simple: our production
is proportional to the $\alpha_Y^2$ rather than $\alpha_2^2 $ which suppresses the cross sections by more than order of magnitude.
On the other hand, if all the new states lie at the same mass scale, we expect the \emph{total} cross sections to be increased by factor of $d_X N_X Q_X^2$.\footnote{The same $d_X N_X Q_X^2$ dependence originates  
from the optical theorem.} Therefore, when  both these factors are taken into account we end up with the cross sections, which are
slightly bigger (by order-one factor) than the standard EW cross sections. 

Most of the searches for the EW states at the LHC for now, are motivated by the SUSY EW-ino. This usually ends up 
in final states leptons (including, possibly, taus) and with $\met$. Although we do not know, what would be exactly 
the bounds on every single scenario one would consider, the direct detection bounds are very unlikely to exceed the bounds on the SUSY wino-chargino particles. The bound on chargino is around 475~GeV if it is assumed to decay to the electrons or muons~\cite{Aad:2014vma}, and it is around 360~GeV if we are considering decays into taus~\cite{Aad:2014yka}.

\section{Conclusions}\label{sec:Conclusions} 

In this paper we made one of the first attempts to confront some theoretical explanations of
the diphoton access with LHC SM precision measurements.
Phenomenological models which try to fit the large width of diphoton excess, suggested by ATLAS, 
suggest large multiplicity of the EW-charged states below the TeV scale that have a significant impact 
on the hypercharge coupling running. We point out that this running can be probed via DY production 
at the LHC, estimate the bounds from LHC8 and project the bounds from the LHC14. We show, that contrary to 
the direct detection our method is robust and it can clearly exclude or confirm such new states at the EW scale. 

Interestingly the bounds that we derive from the LHC8, although still relatively weak, are already much stronger 
that the bounds one get from LEP. Moreover, LHC14 DY measurements will significantly improve the reach, 
pushing the bounds (or maybe making discoveries) deep into the parameter space relevant for the wide 750~GeV 
resonance. 

We also briefly comment on the possibilities of the direct detection of the new EW states at the LHC. Unfortunately 
this question is much more model dependent, and lacks the robustness of the precision measurement approach. 
While it is relatively easy to hide the new states from the direct detection by assuming complicated decay chains and 
large multiplicity state, it would be interesting to survey more carefully various decay modes.
 
\acknowledgments{
The research of FG is supported by a Marie Curie Intra European Fellowship within the 7th European Community Framework Programme (grant no. PIEF-GA-2013-628224).
We thank G.~F.~Giudice, M.~Mangano and 
A.~Strumia for useful discussions. We are also grateful to D.~Bourilkov for correspondence. 
MS is also grateful to 
CERN for hospitality.  
}

\vspace{0.7cm}
\noindent 
{\bf Note added.} When our manuscript was in the final stages of preparation, 
a work of~\cite{Gross:2016ioi} appeared,
which has a substantial overlap
with our work. 
Note however, that our constraints are significantly milder than those claimed by~\cite{Gross:2016ioi}. The 
discrepancy is probably due to different treatment of various  systematic errors and correlations between them. 

\begin{appendix}

\section{Simulating the theoretical prediction for the DY at the LHC}\label{sec:AppA}

\end{appendix}

We describe the theory framework used to derive
the predictions for the DY cross section 
$d\sigma_{\rm DY}/d m \equiv d\sigma(p p \to Z,\gamma \to \ell^+ \ell^-)/d m_{\ell^+\ell^-},\ \ell=e \text{ or } \mu$, 
at the LHC, binned in $m_{\ell^+\ell^-}$ as described in~\cite{hepdata_DY}.
We employ the code {\tt FEWZ}~\cite{Gavin:2010az}, version 3.1b2, which includes NNLO
QCD and NLO EW corrections to the process. The cross section is considered over the full
phase space, {\it i.e.}, no cuts are applied - beyond the standard $p_T>10\,$GeV 
($p_T>20\,$GeV, $\eta<4.5$) cuts on real photons (jets). We use {\tt CTEQ12NNLO} pdfs and 
chose a dynamical scale of $\mu_R=\mu_F=m_{\ell^+\ell^-}$. 

\begin{figure}[!t]
	\begin{center}
	\includegraphics[height=1.65in]{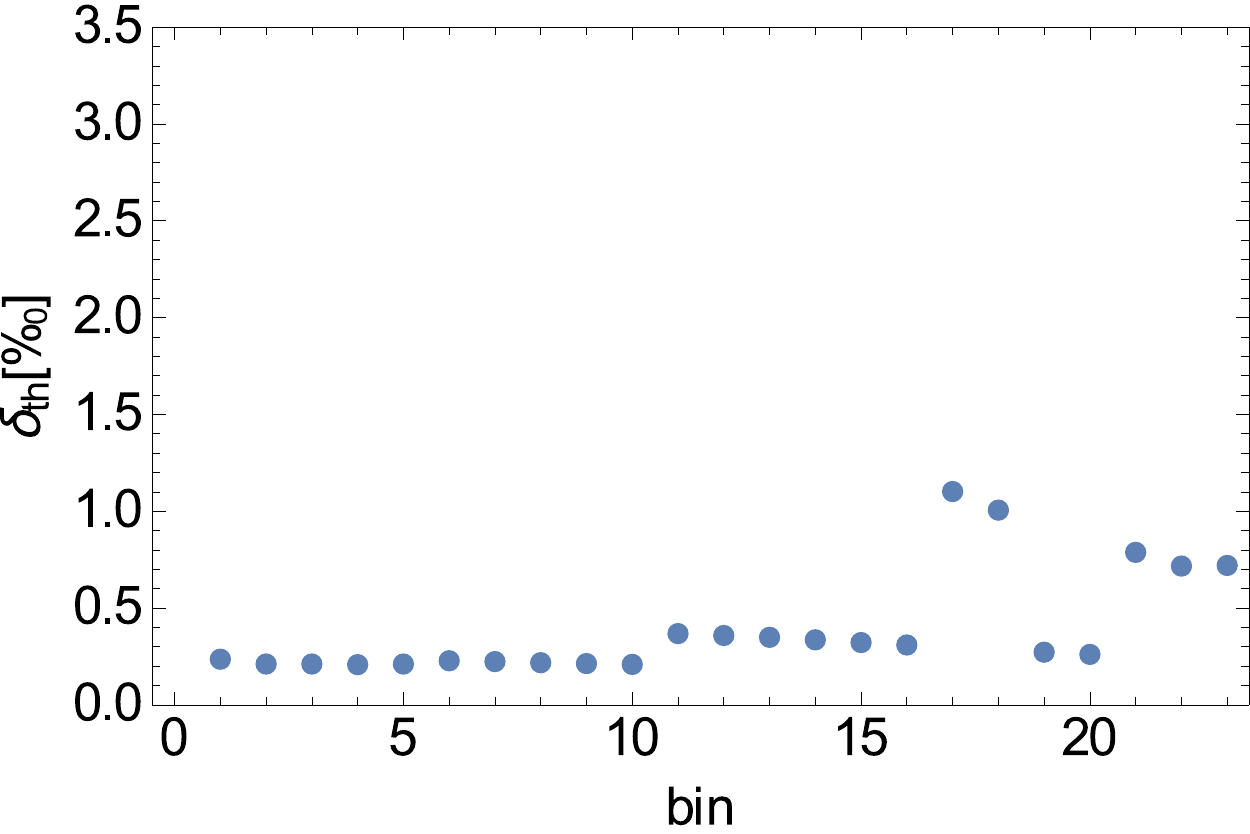} \quad \includegraphics[height=1.65in]{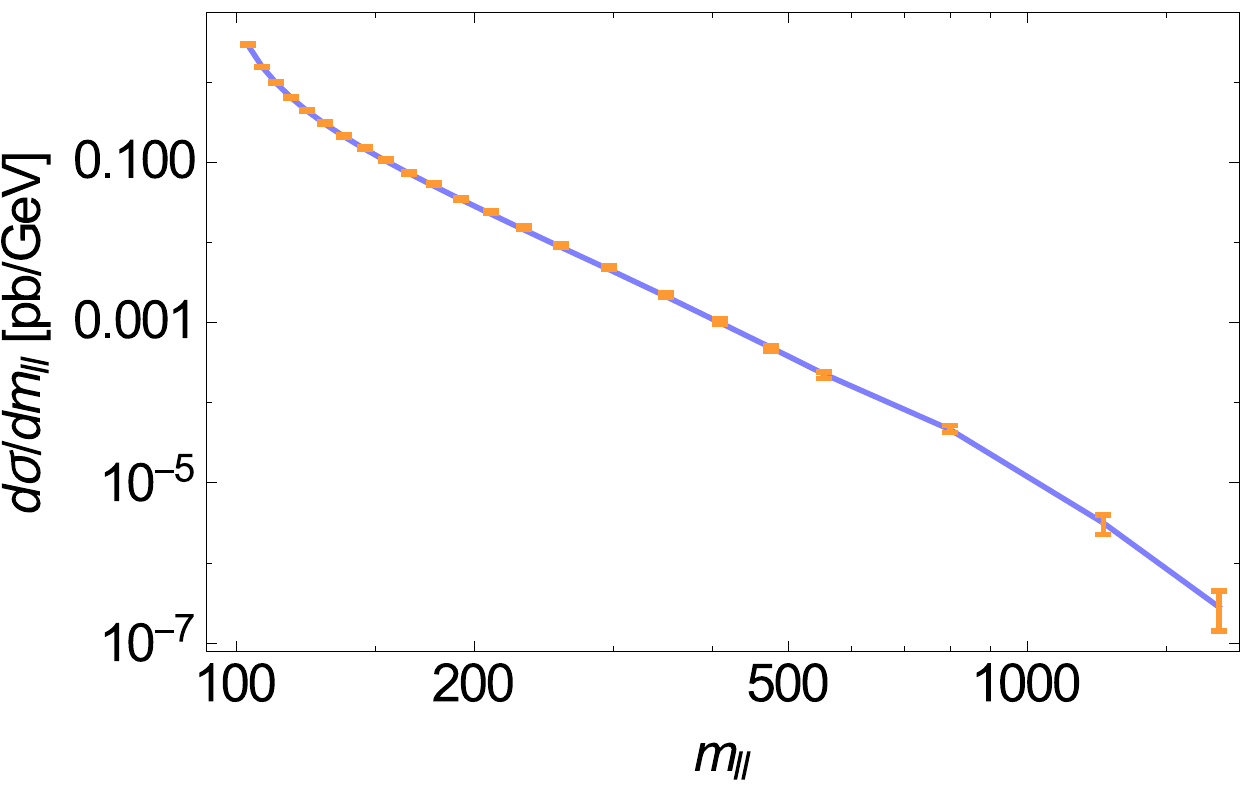}
	\caption{\label{fig:plot} {\it Left: Monte Carlo error quoted by {\tt FEWZ} for each bin in per mille. Right: 
        Theoretical prediction for the cross section $d\sigma_{\rm DY}/d m$ (blue line), overlayed with the 
        corresponding experimental results (orange error bars), at $\sqrt s = 8\,$ TeV. See text for details.}}
	\end{center}
	\end{figure}

	\begin{figure}[!t]
	\begin{center}
	\includegraphics[height=1.65in]{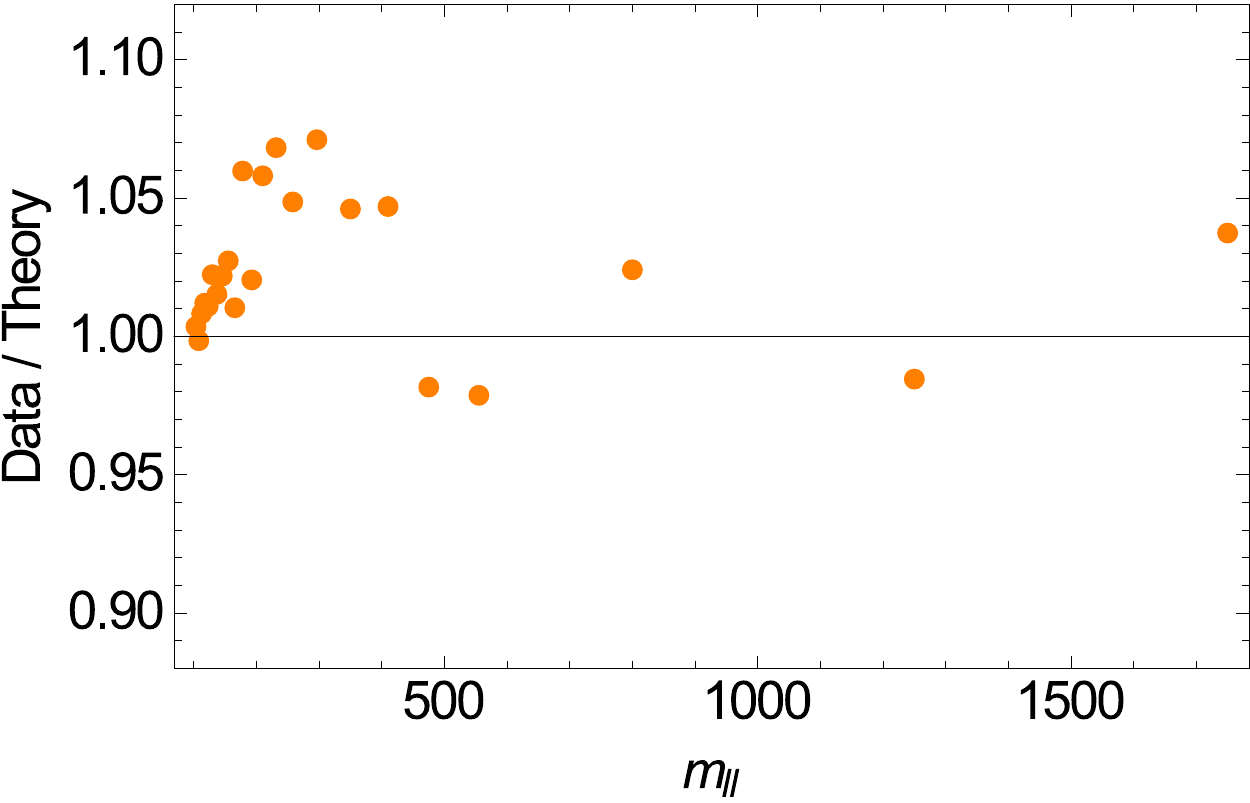} \quad \includegraphics[height=1.65in]{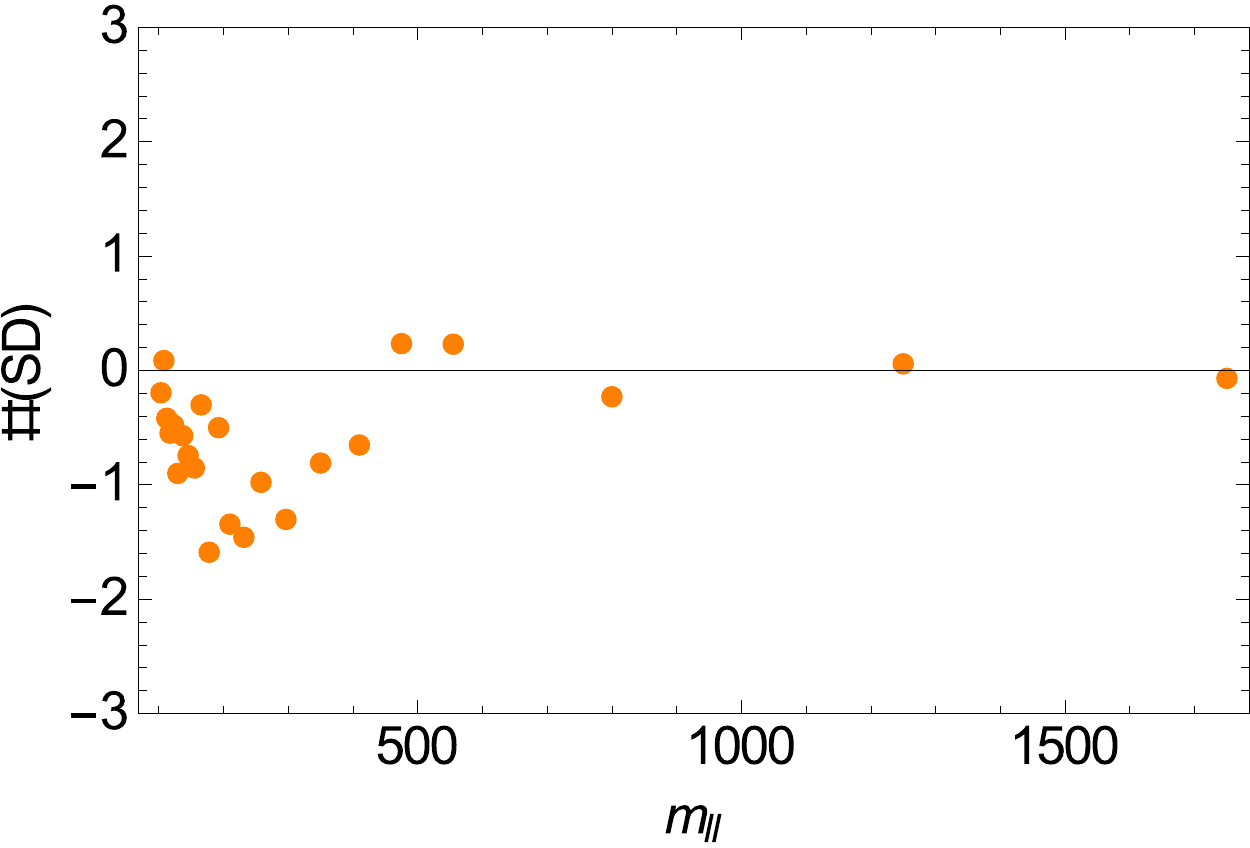}
	\caption{\label{fig:plot2} {\it Left: Ratio $\sigma_{\rm DY}^{\rm th}/\sigma_{\rm DY}^{\rm exp}$.
	Right: Difference between theory prediction and experimental results in standard deviations.
	See text for details.}}
	\end{center}
\end{figure}
	
As CMS presents their data~\cite{hepdata_DY} after unfolding actually both initial state and final state radiation (employing Monte Carlo simulation), we set {\tt EW control = 7} in {\tt FEWZ}.\footnote{D.~Bourilkov, private communication.} This removes the main source of difference between final state electrons and muons.

Moreover, we turn off the photon-induced channel via setting \mbox{\tt Alpha\,QED(0)=0}, 
since it has been removed in the results as given in \cite{hepdata_DY,CMS:2014jea}.
The other physical parameters are kept as in the {\tt FEWZ} default, in particular we employ the
\mbox{$G_\mu$ input} scheme and use the $Z-$pole focus.

\begin{table}[!htb!]
\centering
\begin{tabular}{||c|c|c||}
\hline\hline
$m_{\ell^+\ell^-}$\,[GeV] & $d \sigma_{\rm DY}/d m$\,[pb/GeV] & $\delta_{\rm th}$\,[pb/GeV] \\
\hline\hline
103.5 & 2.95064& 0.000695558 \\  \hline
108 & 1.55449 & 0.00032792 \\  \hline
112.5 & 0.975046 & 0.000205666 \\ \hline
117.5 & 0.641322 & 0.000133134 \\ \hline
123 & 0.440523 & 0.0000928767 \\ \hline
129.5 & 0.303344 & 0.0000691137 \\ \hline
137 & 0.210305 & 0.0000469974 \\ \hline
145.5 & 0.1469 & 0.0000319788 \\ \hline
155 & 0.103285 & 0.0000220165 \\ \hline
165.5 & 0.0730464 & 0.000015263 \\ \hline
178 & 0.0505791 & 0.0000186166 \\ \hline
192.5 & 0.0343989 & 0.0000123249 \\ \hline
210 & 0.0227222 & $7.9255\times10^{-6}$ \\ \hline
231.5 & 0.0143423 & $4.81843\times10^{-6}$ \\ \hline
258 & 0.0086692 & $2.78966\times10^{-6}$ \\ \hline
296.5 & 0.00458432 & $1.41792\times10^{-6}$ \\ \hline
350 & 0.00211278 & $2.3279\times10^{-6}$ \\ \hline
410 & 0.000987678 & $9.93302\times10^{-7}$ \\ \hline
475 & 0.000484899 & $1.31907\times10^{-7}$ \\ \hline
555 & 0.000225813 & $5.88447\times10^{-8}$ \\ \hline
800 & 0.0000459948 & $3.62253\times10^{-8}$ \\ \hline
1250 & $3.13852\times10^{-6}$ &  $2.24954\times10^{-9}$ \\ \hline 
1750 & $2.79568\times10^{-7}$ & $2.01314\times10^{-10}$\\ 
\hline\hline
\end{tabular}
\caption{\label{tab:FEWZ} {\it Theory prediction for the DY cross section at $\sqrt s = 8\,$ TeV together
with the Monte Carlo error. The central values of the bins in $m_{\ell^+\ell^-}$ are
quoted in the first column. See text for details.}}
\end{table}

It turns out that the accuracy of the results in the large $m_{\ell^+\ell^-}$ bins
could be improved by running each bin individually - allowing to generate sufficient 
Monte Carlo statistic also for high-mass bins (which have a limited impact on the total cross section).
We show the resulting errors quoted by {\tt FEWZ} (not including pdf uncertainties or scale variation) 
for $\sqrt s = 8\,$ TeV in the left panel of Figure~\ref{fig:plot} -- demonstrating that 
the Monte Carlo error seems under good control in all bins, {\it i.e.}, 
\beq
\delta_{\rm th} \equiv \delta \sigma_{\rm DY}^{\rm th}/\sigma_{\rm DY}^{\rm th} \lesssim 1 \permil~.
\eeq

Finally, we present our $\sqrt s = 8\,$ TeV results for the differential cross section
in Table~\ref{tab:FEWZ}, including the {\tt FEWZ} errors.
We also provide in the right panel of Figure \ref{fig:plot} a simultaneous plot of 
our predictions (blue, joined - neglecting the small $\delta_{\rm th}$) and the experimental results as
given in~\cite{hepdata_DY} (orange - including the quoted error bars). Moreover, in the left and
right panels of Figure \ref{fig:plot2}, we present for completeness 
the ratio $\sigma_{\rm DY}^{\rm th}/\sigma_{\rm DY}^{\rm exp}$ and the difference between theory and experiment
in standard deviations, adding the corresponding errors in quadrature.

\vspace{1.0cm}



  
  
  

\bibliography{refs}
\bibliographystyle{jhep}
\end{document}